\documentclass[a4paper,11pt]{article}
\pdfoutput=1 

\usepackage{jcappub} 

\usepackage[T1]{fontenc} 
\usepackage{amsmath}
\usepackage{calc}
\usepackage{xcolor}

\usepackage{accents}

\newcommand{\dbtilde}[1]{\accentset{\approx}{#1}}

\subheader{PI/UAN-2021-687FT}

\title{\boldmath Anisotropic Scalar Field Dark Energy with a Disformally Coupled Yang-Mills Field}

\author[a,1]{L.~Gabriel G\'omez,\note{Corresponding author.}}
\author[b,c]{Yeinzon Rodr\'iguez}
\author[d]{Juan P. Beltr\'an Almeida}

\affiliation[a]{Departamento de F\'isica, Universidad de Santiago de Chile,\\Avenida V\'ictor Jara 3493, Estaci\'on Central, 9170124, Santiago, Chile}
\affiliation[b]{Centro de Investigaciones en Ciencias B\'asicas y Aplicadas, Universidad Antonio Nari\~no,\\
Cra 3 Este \# 47A-15, Bogot\'a D.C. 110231, Colombia}
\affiliation[c]{Escuela de F\'isica, Universidad Industrial de Santander,\\ Ciudad Universitaria, Bucaramanga 680002, Colombia}
\affiliation[d]{Departamento de F\'isica, Universidad Nacional de Colombia,\\
Cra 30 \# 45-03, Bogot\'a D.C. 111321, Colombia}

\emailAdd{gabriel.gomez.d@usach.cl}
\emailAdd{yeinzon.rodriguez@uan.edu.co}
\emailAdd{jubeltrana@unal.edu.co}

\abstract{In the context of scalar-tensor theories, the inclusion of new degrees of freedom coupled non-minimally to the gravitational sector might produce some appealing effects on the cosmic expansion history.
We investigate this premise by 
including a canonical SU(2) Yang-Mills field to the total energy budget of the universe coupled to the standard quintessential field by a disformal transformation. 
From the dynamical system analysis, we study three cases of cosmological interest that span most of the physical phase space of the model: the uncoupled limit, the isotropic, and the Abelian cases. New scaling solutions with a non-vanishing gauge field are found in all cases which can be interesting for early cosmological scenarios.  Some of these scaling solutions even exhibit anisotropic features. 
Also, the background evolution of the universe is studied by means of numerical analysis. As an interesting result, the disformal coupling changes the equation of state of the gauge field from radiation to matter at some stages of the evolution of the universe, thereby the gauge field can contribute to some fraction of the total dark matter. We have also quantified the redshift-dependent contribution of the gauge field in the form of dark radiation during the radiation era to the effective number of relativistic species. This depends essentially on the initial conditions and, more importantly, on the disformal coupling function.}

\begin{document}
\maketitle
\flushbottom

\section{Introduction}
\label{sec:intro}
Today we have robust and convincing evidence coming from diverse high precision cosmological observations confirming the current accelerated expansion of the universe \cite{Riess:1998cb,Perlmutter:1998np,Aghanim:2018eyx}. Although those observations had improved and broadened our current understanding of the dynamics of the universe to an unprecedented level of detail, we are still lacking from a fundamental explanation for the agent driving the current accelerated expansion, usually attributed to an exotic fluid component dubbed dark energy (DE).  A cosmological constant term is arguably the simplest candidate for the agent behind the accelerated expansion, but, from a particle physics point of view, this term originates from the zero-point contributions to vacuum fluctuations whose estimate leads to the disastrous discrepancy of around 120 orders of magnitude when compared with its observed value \cite{Amendola:2015ksp}. This dramatic discrepancy constitutes the celebrated cosmological constant problem \cite{RevModPhys.61.1}. In order to evade this problem,  several mechanisms have been proposed to give an alternative explanation for the accelerated expansion, mechanisms in which the cosmological constant is assumed to vanish or becomes negligible compared to present cosmological energy density. Many of those mechanisms rely on dynamical dark energy models traditionally referred as quintessence models. In these models, the quintessence field is a canonical scalar field  slowly rolling along a potential which produces a dynamical equation of state able to mimic the behaviour of a cosmological constant \cite{Amendola:2015ksp}. Naturally, unless some symmetry principle is prescribed, this scalar field is also expected to interact with other components of the universe such as standard model particles, dark matter and additional degrees of freedom \cite{Amendola:1999er}. These interacting scenarios lead to a rich variety of phenomenological consequences which have been studied with great interest over the last decades paying particular attention to the solution of some existing tensions in the $\Lambda$CDM model when confronting with different redshift observations. (see \cite{Mangano:2002gg,Amendola:2003wa,Koivisto:2005nr,Guo:2007zk, Quartin:2008px,Quercellini:2008vh,Pettorino:2008ez,Gannouji:2010fc,Pourtsidou:2013nha,Amendola:2014kwa} for a sample of references).

A particular type of additional component to include in a cosmological setup is a spin-1 vector field. 
Models involving vector or gauge fields have been widely studied over the years and the interest on such models has a long standing history in cosmological contexts \cite{Ford:1989me, Jacobson:2000xp,Dimopoulos:2006ms, Ackerman:2007nb,Dimopoulos:2008yv,Golovnev:2008cf,EspositoFarese:2009aj,Dimopoulos:2011ws}. Studies about vector fields cover different aspects ranging from fundamental theoretical aspects to the calculation/estimation of diverse features and signatures imprinted by the presence of such fields (for reviews on the subject see \cite{Maleknejad:2012fw,Soda:2012zm,Dimastrogiovanni:2010sm} and references therein).  The usefulness of vector fields are traditionally constrained in cosmological setups by the cosmic no-hair theorem \cite{Starobinsky:1982mr,Wald:1983ky} which predicts an exponentially rapid dilution of any anisotropic effects in an accelerated expanding background if there are no anisotropic sources such as vector or gauge fields. Nevertheless, the presence of a non minimal coupling to gravity \cite{Golovnev:2008cf,Kanno:2008gn} and field dependent couplings between the scalar and vector sectors \cite{Watanabe:2009ct,Dimopoulos:2009am} can evade the conditions of the no-hair theorem \cite{Maleknejad:2012as} and allow for small and non-diluting (and eventually observable) features sourced by the vector fields while sustaining a well behaved dynamics with interesting phenomenological features \cite{Dimopoulos:2008yv,Karciauskas:2008bc,Rodriguez:2013cj}. 
Although many studies have focused on the role of vector fields during the inflationary period \cite{Garnica:2021fuu,Emami:2016ldl,Maleknejad:2012fw,Soda:2012zm,Dimastrogiovanni:2010sm,Bamba:2008xa}, some models have also focused on their role  during the late-time evolution of the universe \cite{Bamba:2008xa,Thorsrud:2012mu,DeFelice:2016cri,DeFelice:2016uil,Almeida:2018fwe,Rodriguez:2017wkg,Guarnizo:2019mwf, Guarnizo:2020pkj,Nakamura:2019phn} and even considered vector fields as viable and well motivated dark energy candidates \cite{ArmendarizPicon:2004pm, Koivisto:2007bp, Koivisto:2008ig,Tasinato:2014eka, Mehrabi:2015lfa,DeFelice:2016yws, Alvarez:2019ues, Nakamura:2019phn, Motoa-Manzano:2020mwe,Orjuela-Quintana:2020klr,Gomez:2020sfz,Gomez:2022okq} leading to interesting theoretical and phenomenological consequences.

Motivated by the study of generic scenarios involving the scalar quintessence field and gauge fields and their cosmological consequences on the evolution of the universe at different epochs, in this paper we consider an interacting scenario allowing for couplings between these two sectors of the model.
A particular way of introducing non trivial interactions between the scalar field and any other component is by introducing field dependent metric transformations. An important example of those transformations is the so called {\it disformal transformation}. 
%
%
This type of metric transformations were first introduced  by Bekenstein in \cite{Bekenstein:1992pj} while studying generalized formulations of gravity and its interaction with matter in the presence of two geometrical metrics. Later, it was realized that disformal transformations play a central role in the study of the structure of generalized scalar-tensor theories involving non minimal couplings between gravity and a scalar field \cite{Bettoni:2013diz,Achour:2016rkg}. Those studies on generalized scalar-tensor models revealed deep connections and equivalences between different approaches to modified gravity models such as the Horndeski theory, higher-order curvature and $f(R)$ models and conduced to recent important developments in the understanding of the theory of gravity and its interaction with matter \cite{Zumalacarregui:2013pma,Gleyzes:2014qga,deRham:2010ik,deRham:2010kj}. 

On the other hand, from a more phenomenological point of view, disformal transformations have been used as a straightforward and natural source for non trivial interactions between different sectors of a model. Particularly, previous studies involving disformal couplings in cosmology include \cite{Kaloper:2003yf,vandeBruck:2015tna} for the inflationary analysis, \cite{Koivisto:2008ak,Zumalacarregui:2010wj,Koivisto:2013fta,KOIVISTO:2013jwa,Teixeira:2019hil,vandeBruck:2020fjo} for dark energy and the late time evolution of the universe, and \cite{Cembranos:2016jun} for dark matter models involving disformally coupled scalars of relevance in cosmology. More specifically, cosmological models involving scalar and gauge fields disformally coupled include \cite{KOIVISTO:2013jwa, Koivisto:2014gia,Koivisto:2015vda,Gomez:2022okq}. It is this more phenomenological approach which serves as a motivation for this paper in which we consider a scenario where the quintessence field is coupled via disformal transformations to a $\rm{SU(2)}$ Yang-Mills field while keeping a minimal coupling to matter. This possibility retains the good features of quintessential models and opens a window for interesting phenomenological consequences offered by a non-diluting gauge field. In this work we do not identify the $\rm{SU(2)}$ field with the $W_{\mu}$-boson gauge field or any other field of the standard model of particle physics since the field dependent couplings would lead to a dynamical behaviour of the coupling constants very difficult to make compatible with current observations. Instead, we consider that the gauge field constitutes part of a dark sector which does not interact directly with other matter fields but does it indirectly due to the mediation of the quintessence field and gravity. 

Due to the introduction of novel couplings in the model and the presence of matter fields, the model considered here exhibits a rich variety of possibilities in its parameters and phase space similar to coupled quintessence models. For instance, this model features new scaling solutions 
related to the presence of gauge field hair, during radiation, matter and dark energy dominated periods. In particular, the presence of gauge hair, leads to new types of anisotropic scaling solutions. This anisotropic hair could be relevant, for instance, to give some insight about the origin of the cosmic microwave background (CMB) quadrupole anomaly \cite{Aghanim:2018eyx, Appleby:2009za,Koivisto:2008ig}.  Another remarkable feature of the model is a non-standard, field dependent equation of state for the ${\rm SU(2)}$ field. The disformal coupling changes the equation of state of the gauge field, making this component perform differently from radiation as in standard minimally or conformally coupled case in which $\omega_A = 1/3$. This novel dynamical behaviour offers new possibilities for the evolution of the ${\rm SU(2)}$ field so that it can act as radiation at some stages and transit to a dark matter component with $\omega_A=0$ at some other stages. This characteristic of the disformally coupled gauge field is one of the main properties that we are going to exploit in the dynamical analysis of the system supported by numerical analysis. Additionally, the inclusion of gauge ${\rm SU(2)}$ fields, increases the number of relativistic degrees of freedom with potential important effects for the cosmic expansion rate, CMB anisotropies and the distribution of large-scale structures (LLSs). 

This paper is organized as follows. In section \ref{sec:II} we describe the basics of the model and introduce the disformal transformations relating the scalar and gauge field and derive the field equations of the system. In order to analyse the dynamical evolution of the model,  we describe the autonomous system in a Bianchi-I background metric in section \ref{sec:III} featuring the anisotropic nature of the model. In section \ref{sec:IV} we present a detailed description of the phase space, identifying relevant fixed points and interesting regions in the parameter space, and discuss its relevance in the context of the late-time evolution of the universe. We also support this description on a numerical analysis of the background evolution. In section \ref{conclusions} we summarize and discuss our main results.

Throughout the text we will use Greek indices to denote space-time indices running from 0 to 3 while Latin indices are internal $\rm{SU(2)}$ group indices running from 1 to 3. The sign convention is the (+++) according to Misner, Thorne, and Wheeler \cite{Misner:1974qy}.

\section{The model}
\label{sec:II}

We start by describing the total action in the Einstein frame consisting of a scalar tensor theory, the standard matter  sector (composed of dark matter, baryons and radiation), and a SU(2) Yang-Mills field:
\begin{equation}
S=\int d^{4} x \sqrt{-g}\left[\frac{M_{\rm p}^{2}}{2}R+p(\phi, X)+\mathcal{L}_{M}(g_{\mu\nu},\Psi_{i})\right]-\int d^{4} x \sqrt{-\hat{g}}\left[\frac{1}{4} \hat{g}^{\alpha \beta} \hat{g}^{\gamma \delta} F^{a}_{ \alpha \gamma} F_{\beta  \delta a}\right] .\label{eq1}
\end{equation}
In this expression, $g$ is the determinant of the gravitational metric $g_{\mu \nu}$, $M_p$ is the reduced Planck mass, $R$ is the Ricci scalar, $p(\phi, X)$ is an arbitrary function of the scalar field $\phi$ and its kinetic term $X\equiv-\frac{1}{2}g^{\mu\nu}\phi_{, \mu} \phi_{, \nu}$, $\mathcal{L}_M$ is the matter Lagrangian where $\Psi_{i}$ denotes the matter fields, $\hat{g}$ is the determinant of the (disformal) metric $\hat{g}_{\mu \nu}$ felt by the Yang-Mills field $A_\mu$, and $F_{\mu\nu}^{a}$ is the gauge-field strength tensor defined as $F_{\mu\nu}^{a}\equiv A_{\nu,\mu}^{a}-A_{\mu,\nu}^{a}+\tilde{g} \epsilon^{a}_{\;\; cd} A_{\mu}^{c} A_{\nu}^{d}$, $\tilde{g}$ being the group coupling constant and $\epsilon^{a}_{\;\; cd}$ being the Levi-Civita symbol which represents the structure constants of the SU(2) group. The action for the Yang-Mills field
can be rewritten as 
\begin{equation}S_{A}=\int d^{4} x \sqrt{-g}\left[\gamma^{-1}\left(-\frac{1}{4} F^{2}\right)+\frac{1}{2} \gamma h \phi^{,\alpha} \phi^{,\beta}\left(F^{a}
_{\gamma \alpha} F_{\;\; \beta a}^{\gamma}\right)\right],\label{eq2}
\end{equation}
by making use of the disformal transformation of the form
\begin{equation}
\hat{g}_{\mu \nu}=C(\phi) g_{\mu \nu}+D(\phi) \phi_{, \mu} \phi_{, \nu} \,,\label{eq3}
\end{equation}
which relates the geometries felt by the the gauge field and the scalar field respectively (the metric $\hat{g}_{\mu\nu}$ is associated to the gauge field sector of the model, meaning that the gauge field propagates along $\hat{g}_{\mu\nu}$ geodesics, while the scalar field propagates following ${g}_{\mu\nu}$ geodesics). Here $C(\phi)$ and $D(\phi)$ are, as usual, the conformal and disformal couplings, respectively, and the coupling factor in Eqn. (\ref{eq2}) is defined as
\begin{equation}
\gamma \equiv\left(1-2 h X\right)^{-\frac{1}{2}} \,,\label{eq4}
\end{equation}
with $h \equiv D/C$. 

It is interesting to note that the Yang-Mills action is conformally invariant since the factors $C^{2}$ coming from the Jacobian of the transformation of the metric and the inverse of the disformal metric cancel out so that no term in the action gets rescaled by $C$. Hence, the purely conformal case (with $C$ arbitrary), $D=0$ ($h=0$) and $\gamma=1$, coincides with the uncoupled limit of the model. The same conclusion was attained in the disformally coupled Abelian vector model \cite{Koivisto:2014gia} once the mass term is removed. Notice also that when the velocity of the scalar field vanishes, $\gamma\to1$, but the action Eqn.~(\ref{eq2}) is still affected by $h$. So this latter possibility does not lead successfully to the uncoupled case. With all this, it is plain to notice the appearance of the derivative coupling of the scalar field to the gauge field in $S_{A}$ (Eqn.~(\ref{eq2})) in the disformal coupling case. We could also have included the kinetic term $X$ dependence in the conformal and disformal coupling functions but it would have complicated the system tremendously, so we have discarded this possibility.\footnote{Additionally, the dependence of the scalar kinetic term $X$ would imply problems with the Hamiltonian stability of the model and with the hyperbolicity of the scalar equation of motion in certain domains of the parameter space \cite{Zumalacarregui:2013pma,Fleury:2014qfa}. So, we will restrict to the $C(\phi)$ and $D(\phi)$ case here only.} For simplicity, we have considered the canonical scalar field Lagrangian as $p(\phi, X)=X-V(\phi)$. So, the novelty of our proposal relies on the fact that the scalar field is, in general, uncoupled from the matter sector, i.e., this part of the action looks like similarly in both the Einstein and Jordan frames, but a new sector given by the gauge field is coupled to the scalar field Lagrangian\footnote{There also exists a frame which the gauge field is minimally coupled in, but the price to pay in this frame is that the remaining sectors, including the gravitational one, exhibit explicit (derivative) interactions with the scalar field.}.  
Hence, this novel coupling affects the evolution of the scalar field itself and allows the gauge field to evolve jointly with it during the cosmological evolution, contributing thus to the total energy budget of the universe at different epochs. This intriguing  aspect as well as the derived phenomenological consequences are our primary concerns in this work.

We remind the reader that coupling the matter sector to the scalar-tensor sector so that both matter and scalar field propagate in different geodesics is what is most commonly found in the literature. Such a scenario, 
\cite{Amendola:1999er,Amendola:1999qq,Koivisto:2008ak,vandeBruck:2015ida,Zumalacarregui:2010wj,Zumalacarregui:2012us,vandeBruck:2013yxa,vandeBruck:2016jgg,Teixeira:2019hil,Teixeira:2019tfi}, has been categorized as a coupled dark energy model.  Within this scenario, another phenomenological possibility is to add a Maxwell-type vector field coupled to  a quintessence field as well as to dark matter
 \cite{Thorsrud:2012mu}. To clarify things, we shall refer henceforth to the uncoupled scalar field as the one in the $\phi$CDM reference model which is the standard quintessence scenario \cite{Peebles:1987ek,Wetterich:1987fm}.

Varying the action in Eqn.~(\ref{eq1}) (considering also Eqn. (\ref{eq2})) with respect to $g_{\mu\nu}$ leads to the field equations     
\begin{equation}
M_{\rm p}^{2} G_{\mu \nu}=T_{\mu \nu}^{\phi}+T_{\mu \nu}^{A}+T_{\mu \nu}^{M} \,,\label{eq5}
\end{equation}
where $G_{\mu \nu}$ is the Einstein tensor, $T_{\mu \nu}^{\phi}$ is the energy-momentum tensor of the scalar field:
\begin{equation}
T_{\mu \nu}^{\phi}=-\frac{1}{2}\left(g^{\alpha\beta}\phi_{, \alpha}\phi_{, \beta} +V(\phi)\right)g_{\mu\nu}+\phi_{, \mu}\phi_{, \nu} \,,\label{eq6}
\end{equation}
$T_{\mu\nu}^{A}$ is the effective energy-momentum tensor of the gauge field:
\begin{align}
T_{\mu\nu}^{A}=&\frac{F_{\mu}^{\;\; \alpha a} F_{\nu\alpha a}}{2 \gamma}-\frac{1}{2} h F^{a}_{\mu\alpha} F_{\nu\beta a} \gamma \nabla^{\alpha} \phi \nabla^{\beta} \phi+g_{\mu\nu}\left(-\frac{F^a_{\alpha\beta} F^{\alpha \beta}_a}{8\gamma}+\frac{1}{4} h F_{\alpha \;\; a}^{\;\; \eta} F_{\beta\eta}^a \gamma \nabla^{\alpha} \phi \nabla^{\beta} \phi\right) \nonumber \\
& -\frac{1}{2} h F_{\alpha\beta}^a F_{\nu \;\; a}^{\;\; \beta} \gamma \nabla^{\alpha} \phi \nabla_{\mu} \phi-\frac{1}{2} h F_{\alpha\beta}^a F_{\mu \;\, a}^{\;\; \beta} \gamma \nabla^{\alpha} \phi \nabla_{\nu} \phi+\frac{1}{8} h F_{\alpha\beta}^a F^{\alpha\beta}_a \gamma \nabla_{\mu} \phi \nabla_{\nu} \phi \nonumber \\
& + \frac{1}{4} h^{2} F_{\alpha \;\; a}^{\;\; \eta} F_{\beta\eta}^a \gamma^{3} \nabla^{\alpha} \phi \nabla^{\beta} \phi \nabla_{\mu} \phi \nabla_{\nu} \phi \,,\label{eq7}
\end{align}
and $T_{\mu\nu}^{M}$ is the energy-momentum tensor of matter: 
\begin{equation}
    T_{\mu\nu}^{M}\equiv  -\frac{2}{\sqrt{-g}}\frac{\delta \mathcal{S}_{M}}{\delta g^{\mu\nu}} \,,\label{eq7a}
 \end{equation}
where $\mathcal{S}_{M}$ is defined, in turn, as the third term in Eqn.~(\ref{eq1}). Now, varying the action with respect to the scalar and gauge fields leads to the field equations
\begin{equation}
\square \phi-V_{\,\phi}=Q \,,\label{eq8}
\end{equation}
and 
\begin{equation}
D_{\mu}\mathcal{F}^{\mu\nu b}=0 \,,\label{eq9}
\end{equation}
where $\mathcal{F}^{\mu\nu b} \equiv \frac{F^{\mu\nu b}}{2\gamma}-h\gamma F^{\;\;\mu b}_{\alpha}\nabla^{\nu} \phi \nabla^{\alpha} \phi$, $D_{\mu} \equiv \nabla_{\mu}-i \tilde{g}\left[A_{\mu}, *\right]$ denotes the gauge covariant derivative, 
and $Q$ denotes the interacting term which acquires the explicit form 
\begin{equation}Q=\gamma^{2}\left[T^{A \mu \nu} \nabla_{\mu}\left(\frac{D}{C} \phi_{, \nu}\right)-\frac{1}{2 C}\left(C_{,\phi} T^{A}+D_{,\phi} T^{A \mu \nu} \phi_{, \mu} \phi_{, \nu}\right)\right] \,,\label{eq11}
\end{equation}
with $T^{A}$ being the trace of $T^{A}_{\mu\nu}$.
Thus, the total energy-momentum tensor is covariantly conserved with matter fields being uncoupled:
\begin{equation}
\nabla^{\mu} T_{\mu \nu}^{\phi}=Q \phi_{, \nu} \,, \quad \nabla^{\mu} T_{\mu \nu}^{A}=-Q \phi_{, \nu} \,,\quad \nabla^{\mu} T_{\mu \nu}^{M}=0 \,,\label{eq10}
\end{equation}
where $T^{A}$ is the trace of $T^{A}_{\mu\nu}$. Once the disformal coupling is turned off, this is $h=0$ and, therefore, $\gamma=1$, the standard Yang-Mills field and the canonical quintessence field become fully uncoupled as can be easily checked.

\section{Background cosmology}
\label{sec:III}
The presence of a vector field, which in general will have spatial components,  inhibits the isotropic expansion of the universe.  This happens unless some specific configuration is invoked \cite{Witten:1976ck,Sivers:1986kq,Forgacs:1979zs} such as the temporal gauge \cite{DeFelice:2016yws} or the cosmic triad \cite{Garnica:2021fuu,ArmendarizPicon:2004pm,Maleknejad:2011jw,Nieto:2016gnp,Alvarez:2019ues,Gomez:2020sfz,Rodriguez:2017wkg,Rodriguez:2017ckc}. 
A minor variation of the cosmic triad can, therefore, model an anisotropic expansion still consistent with observations \cite{Maleknejad:2011jr,Murata:2011wv,Orjuela-Quintana:2020klr}.
Accordingly, we shall consider, as a cosmological background spacetime, the axially symmetric Bianchi type-I metric in Cartesian coordinates
\begin{equation}d s^{2}=-d t^{2}+e^{2 \alpha(t)}\left[e^{-4 \sigma(t)} d x^{2}+e^{2 \sigma(t)}\left(d y^{2}+d z^{2}\right)\right] \,,\label{eq12}
\end{equation}
where $e^{\alpha(t)}=a(t)$ is the average scale factor and $e^{\sigma(t)}$ is the spatial shear that characterizes the anisotropy of the expansion.  The global expansion rate (the Hubble parameter) is $H \equiv \dot{\alpha}$ and the anisotropic expansion rate is $\dot{\sigma}$. Hence, to be consistent with the rotational symmetry in the space-like plane orthogonal to the $x$-direction, 
we will consider the following ansatz \cite{Maleknejad:2011jr,Murata:2011wv,Orjuela-Quintana:2020klr}:
\begin{equation}A_{1}^{1}(t) \equiv u(t) \,, \quad A_{2}^{2}(t)=A_{3}^{3}(t) \equiv v(t) \,,\label{eq13}
\end{equation}
while the scalar field takes the simple form $\phi(x^{\mu})=\phi(t)$. Once the space-time geometry and field configurations have been established, the following 
field equations are found:
\begin{eqnarray}
\dot{\alpha}^{2}-\dot{\sigma}^{2}=\frac{1}{3  M_{\rm p}^2}\Big[V(\phi)+\frac{\dot{\phi}^{2}}{2}+\rho_{m}+\rho_{r}
+e^{-4(\alpha+\sigma)} \tilde{g}^{2}v^{2}\frac{\gamma}{2} (2 e^{6\sigma} u^{2}+v^{2}) && \nonumber \\
+e^{-2(\alpha+\sigma)}\frac{\gamma^{3}}{2}(e^{6\sigma} \dot{u}^{2}+2\dot{v}^{2})\Big] &=& 0 \,, \nonumber \\
&& \label{eq14} \\
3(\dot{\alpha}^{2}+\dot{\sigma}^{2})+6\dot{\alpha}\dot{\sigma}+2(\ddot{\alpha}+\ddot{\sigma}) = \frac{1}{M_{\rm p}^2}\Big[V(\phi)-\frac{\dot{\phi}^{2}}{2}+\frac{\rho_{r}}{3}+e^{-4(\alpha+\sigma)} \frac{\tilde{g}^{2}v^{2}}{2\gamma}(-2 e^{6\sigma} u^{2}+v^{2}) && \nonumber \\ 
+e^{-2(\alpha+\sigma)}\frac{\gamma}{2}(e^{6\sigma} \dot{u}^{2}-2\dot{v}^{2})\Big] &=& 0 \,, \nonumber \\
&& \label{eq15} \\
%
3(\dot{\alpha}^{2}+\dot{\sigma}^{2})-3\dot{\alpha}\dot{\sigma}+2\ddot{\alpha}-\ddot{\sigma}= \frac{1}{M_{\rm p}^2}\left[V(\phi)-\frac{\dot{\phi}^{2}}{2}+\frac{\rho_{r}}{3}-e^{-4(\alpha+\sigma)} \frac{\tilde{g}^{2}v^{4}}{2\gamma}- e^{-2\alpha+4\sigma}\frac{\gamma}{2} \dot{u}^{2}\right] &=& 0 \,, \nonumber \\
&& \label{eq16}
\end{eqnarray}
where a dot represents differentiation with respect to the cosmic time $t$ and $\rho_m$ and $\rho_r$ represent the energy densities for matter and radiation respectively whose evolution equations are the usual ones:
\begin{eqnarray}
\dot{\rho}_m + 3 \dot{\alpha} \rho_m &=& 0 \,, \\
\dot{\rho}_r + 4 \dot{\alpha} \rho_r &=& 0 \,.
\end{eqnarray}
Meanwhile, the field equations for the gauge field components are   
\begin{eqnarray}
\ddot{u}+ \dot{u}(\dot{\alpha}+4\dot{\sigma})+2 e^{-2(\alpha+\sigma)}\frac{\tilde{g}^{2}}{\gamma^{2}} u v^{2}+\frac{\gamma^{2}\dot{u}\dot{\phi}}{2}(h_{\phi}\dot{\phi}^{2}+2 h \ddot{\phi}) &=& 0 \,, \label{eq17a} \\
\ddot{v}+\dot{v}(\dot{\alpha}-2\dot{\sigma})+ e^{-2\alpha+4\sigma}\frac{\tilde{g}^{2}}{\gamma^{2}} v u^{2}+\frac{\gamma^{2}\dot{v}\dot{\phi}}{2}(h_{\phi}\dot{\phi}^{2}+2 h \ddot{\phi})+e^{-2(\alpha+\sigma)}\frac{\tilde{g}^{2} v^{3}}{\gamma^{2}} &=& 0 \,,\label{eq17b}
\end{eqnarray}
%
where the subscript $\phi$ means a derivative with respect to the scalar field, which encodes unique information on the coupling functions everywhere in the above and below equations. We will see that this is, in fact, the reason why the assumed phenomenological couplings, discussed in sec. \ref{subsec:4.5} before the conclusions, do not affect the main results we find in terms of $h$.
In the absence of the coupling of the Yang-Mills field to the scalar field, that is, taking properly the limit $\gamma\to1$ or equivalently $h\to0$, Eqs. (\ref{eq17a}) and (\ref{eq17b}) reduce, as expected, to the Yang-Mills equations in an anisotropic Bianchi type-I universe (for example, make the function $f(\phi)=1$ in the anisotropic model described in \cite{Murata:2011wv} for an illustrative comparison). Notice also that these equations are equivalent to the Abelian counterpart in the case $v=0$ \cite{Watanabe:2009ct}. This latter case is an important version of this scenario that we shall consider to check the generality of our results. We lastly comment that the Abelian scenario we will explore in Sec. \ref{subsubsec:4.1.3} was investigated in \cite{Koivisto:2014gia} in the context of inflation with a disformal coupling mediated by a Dirac-Born-Infeld scalar field. Finally, the field equation for the scalar field takes the unfamiliar form

\begin{equation}
\begin{aligned}
&\frac{1}{2}e^{-4(\alpha+\sigma)}\gamma^3 \ddot{\phi}\left(\tilde{g}^2 h v^2 (2 e^{6\sigma}u^2+v^2)+e^{2(\alpha+\sigma)}h \gamma^2(e^{6\sigma}\dot{u}^2+2\dot{v}^2)\right) \\ 
&-\frac{1}{2}e^{-4(\alpha+\sigma)}\gamma \dot{\phi}\Big(2 e^{6\sigma} \tilde{g}^{2} h u^{2}v^{2}(\dot{\alpha}-2\dot{\sigma})+ 2 e^{2(\alpha+\sigma)}h \gamma^2\dot{v}^{2}(\dot{\alpha}-2\dot{\sigma})+ \tilde{g}^{2} h v^{4} (\dot{\alpha}+4\dot{\sigma}) \\
&+e^{2(\alpha+4\sigma)}h \gamma^2\dot{u}^{2}(\dot{\alpha}+4\dot{\sigma})\Big)-\frac{h_{\phi}}{4h^{2}}\Big( e^{-4(\alpha+\sigma)} \tilde{g}^{2} h v^{2}(2e^{6\sigma}u^{2}+v^{2})\gamma (1-\gamma^2)- 2(\gamma^3-3\gamma+2) \\
&+ e^{-2(\alpha+\sigma)} h \gamma^{3} (1-\gamma^{2})(e^{6\sigma}\dot{u}^{2}+2\dot{v}^{2})\Big) \\
&+\gamma^3\ddot{\phi}+3\gamma\dot{\alpha}\dot{\phi}+V_{\phi}=0 \,.\label{eq18}
\end{aligned}
\end{equation}
 It is worthwhile noticing that the last line of Eqn.~(\ref{eq18}) is precisely the canonical Klein-Gordon equation in the case $\gamma=1$. Therefore, it is evident the unpleasant form of this equation and the suggestive necessity of making some helpful assumptions in the dynamical system approach before fully solving it numerically. This issue will be addressed in the following sections.

\subsection{Autonomous system}
\label{sec:IIIa}
In order to figure out the dynamical behaviour of the cosmological model and the stability of the emergent cosmological solutions, we proceed to analyze the system by using the well-known dynamical system techniques (see \cite{Bahamonde:2017ize,Coley:2003mj,Wainwright2009}). Such analysis is roughly based on finding the fixed points and studying their stability under the standard criteria (see e.g. \cite{joseandsaletan}). 
To do so, we first need to define the dimensionless variables
\begin{equation}\begin{aligned}
&x \equiv \frac{\dot{\phi}}{\sqrt{6}M_{\mathrm{P}}H} \,,\quad y \equiv \frac{1}{M_{\mathrm{P}} H} \sqrt{\frac{V}{3}} \,,\quad \Omega_{m}\equiv \frac{\rho_{m}}{3M_{\rm p}^{2}H^{2}} \,, \\ 
& \Omega_{r}\equiv \frac{\rho_{r}}{3M_{\rm p}^{2}H^{2}} \,, \quad p \equiv e^{-\alpha+2\sigma} u \sqrt{\frac{\tilde{g}}{\sqrt{6}m_{\mathrm{P}}H}} \,,\quad s \equiv e^{-(\alpha+\sigma)} v \sqrt{\frac{\tilde{g}}{\sqrt{6}M_{\mathrm{P}}H}} \,, \\
& \Sigma \equiv \frac{\dot{\sigma}}{H} \,, \quad \dbtilde{g} \equiv \left(\frac{3}{2}\right)^{1/4} \sqrt{\frac{\tilde{g} M_{\mathrm{P}}}{H}} \,,\quad w \equiv \frac{e^{-\alpha+2\sigma} \dot{u}}{\sqrt{6}M_{\mathrm{P}} H} \,, \\
& q \equiv \frac{e^{-(\alpha+\sigma)} \dot{v}}{\sqrt{3}M_{\mathrm{P}} H} \,,\label{eq19}
\end{aligned}\end{equation}
where, for convenience, we have normalized them by means of the average expansion rate $H\equiv \dot{\alpha}$. 

The physical meaning of the  dynamical variables is described as follows. To quantify the degree of anisotropy in the expansion rate, we have introduced the dimensionless shear $\Sigma$. The quantities $p$ and $s$ are the components of the non-Abelian field and $w$ and $q$ are their respective velocities. $\dbtilde{g}$ is the normalized version of the $SU(2)$ group coupling constant. The existence of these quantities in the fixed points will indicate a contribution of the gauge field in the form of dark radiation or dark matter at the corresponding epoch as we will see.  The other quantities are the standard ones in the quintessence scenario: $x$ stands for the kinetic energy and $y$ does it for the potential energy of the scalar field.  Overall, we are interested in looking for possible contributions of the new quantities to the resulting fixed points that may describe either a radiation dominated era, a matter dominated era, or drive the current accelerated expansion. So, the existence of the gauge vector field and anisotropy in the fixed points will reveal deviations from the standard quintessence scenario. In doing so, we need to distinguish, as appropriate, three broad cases when classifying fixed points in the cosmological dynamical system: saddle points, unstable points (or repellers) and stable solutions (attractors) \cite{Bahamonde:2017ize}. On the other hand,  Eqn.~(\ref{eq14}) becomes the Friedmann constraint
\begin{equation}
x^{2}+y^{2}+\Omega_{m}+\Omega_{r}+s^{2}\left(2 p^{2}+s^{2}\right) \gamma+\left(q^{2}+w^{2}\right) \gamma^{3}+\Sigma^{2}=1 \,,\label{eq20}
\end{equation}
where the coupling $\gamma$, it being defined according to Eqn.~(\ref{eq4}), appears explicitly as a factor of the energy density associated to the vector field only.  If we define the energy density of dark energy and the vector field respectively as
\begin{equation}\Omega_{\rm DE}=x^{2}+y^{2}+\Sigma^{2}\,, \quad \Omega_{A}=s^{2}\left(2 p^{2}+s^{2}\right) \gamma+\left(q^{2}+w^{2}\right) \gamma^{3}\,,\label{eq21a}
\end{equation}
which will be used later in the numerical computations, the Friedmann constraint can be rewritten simply as 
\begin{equation}
\Omega_{\rm DE}+\Omega_{A}+\Omega_{m}+\Omega_{r}=1 \,.
\end{equation}
Notice that we have found convenient  to identify the Yang-Mills field as an extra dark fluid with density parameter $\Omega_{A}$, and not to taking it in $\Omega_{\rm DE}$, in order to track distinctly its effects in a more clear way during the universe's evolution. Moreover, the energy density of dark energy is defined assuming, in addition to the scalar field contributions, effects coming from the shear. On the other hand,  the model parameters are defined as
\begin{equation}\lambda_{V}=\sqrt{\frac{3}{2}} \frac{V_{,\phi}}{V}M_{\mathrm{P}}\,, \quad \lambda_{h}=\sqrt{\frac{3}{2}} \frac{h_{,\phi}}{h}M_{\mathrm{P}}\,,\label{eq21}
\end{equation}
which allow us to write, along with the dimensionless variables, the coupled system Eqns.~(\ref{eq14})-(\ref{eq18}) in the form of an autonomous system: 
\begin{align}
x^{\prime}&=\Pi+x \epsilon_{H} \,,\\ y^{\prime}&=y\left( \epsilon_{H}+ x \lambda_{V}\right) \,, \\
 \Omega_{m}^{\prime}&= \Omega_{m}\left(2 \epsilon_{H}-3 \right) \,,\\
\dbtilde{g}^{\prime}&=\frac{\dbtilde{g}}{2}\left(1+\epsilon_{H}\right) \,,\\ \gamma^{\prime}&=\gamma\left(\gamma^{2}-1\right)\left(\frac{\Pi}{x}+x \lambda_{h}\right) \,,\\ 
\Sigma^{\prime}&=\left(\frac{\ddot{\sigma}}{\dot{\alpha}^{2}}+\epsilon_{H} \Sigma \right) \,,\\ 
s^{\prime}&=\dbtilde{g}q-s\left(1+\Sigma-\frac{\epsilon_{H}}{2}\right) \,, \\
p^{\prime}&=\sqrt{2} \dbtilde{g} w+\frac{p}{2}\left(-2+4 \Sigma+\epsilon_{H}\right) \,,\\ q^{\prime}&=-\frac{2\dbtilde{g} s\left(p^{2}+s^{2}\right)}{\gamma^{2}}+q\left(-2-\frac{\Pi(\gamma^{2}-1)}{x}+\Sigma+\epsilon_{H}-x\left(\gamma^{2}-1\right) \lambda_{h}\right) \,, \\
w^{\prime}&=-\frac{2\sqrt{2}\dbtilde{g} p s^{2}}{\gamma^{2}}+w\left(-2-\frac{\Pi(\gamma^{2}-1)}{x}-2 \Sigma+\epsilon_{H}-x\left(\gamma^{2}-1\right) \lambda_{h}\right) \,.\label{eq22}
\end{align}
Here, the prime denotes a derivative with respect to $N\equiv\ln{a}$,  $\Pi$ is defined as
%
\begin{equation}
\begin{aligned}
\Pi \equiv &\frac{\ddot{\phi}}{\sqrt{6}M_{\mathrm{P}}H^{2}} \\
=&-\frac{1}{\gamma\left(2x^{2}+\gamma\left(\gamma^{2}-1\right)\left(2 p^{2} s^{2}+s^{4}+\left(q^{2}+w^{2}\right)\gamma^{2}\right)\right)} x [6 x^{2}\gamma+2p^{2}s^{2}\left(\gamma^{2}-1\right)(-1+2\Sigma)- \\
& s^{4}\left(\gamma^{2}-1\right)(1+4\Sigma)+\gamma^{2}\left(\gamma^{2}-1\right)\left(q^{2}(-1+2\Sigma)-w^{2}(1+4\Sigma)\right)+x\gamma(\gamma\left(\gamma^{2}-1\right)(2p^{2}s^{2}+ \\
& s^{4}+\left(q^{2}+w^{2}\right)\gamma^{2})\lambda_{h}+2y^{2}\lambda_{V}) ] \,,\label{eq23} \end{aligned}
\end{equation}
and $\epsilon_{H}$ is given by
%
\begin{eqnarray}
\epsilon_{H} &\equiv& -\frac{\dot{H}}{H^{2}} \nonumber \\
&=& -\frac{2 p^{2} s^{2}(\gamma^{2}-1)+s^{4}(\gamma^{2}-1)+\gamma\left(\Omega_{m}-2x^{2}+4y^{2}+(q^{2}+w^{2})\gamma(\gamma^{2}-1)-2(2+\Sigma^{2})\right)}{2\gamma} \,. \nonumber \\
&& \label{eq24}
\end{eqnarray}
Moreover, by combining Eqns.~(\ref{eq15})-(\ref{eq16}), we find 
\begin{equation}
\frac{\ddot{\sigma}}{\dot{\alpha}^{2}} =\frac{-2 p^{2} s^{2}+2 s^{4}-\left(q^{2}-2 w^{2}\right) \gamma^{2}}{\gamma}-3\Sigma \,.\label{eq25}
\end{equation}
%
%
%
The equation of state of dark energy is then given by
\begin{equation}
 w_{\rm DE}=\frac{x^{2}-y^{2}+\Sigma^{2}}{x^{2}+y^{2}+\Sigma^{2}}\,\label{eq25b},
\end{equation}
and, interestingly, the equation of state of the vector field acquires a non-standard form because of the disformal coupling $\gamma$:
\begin{equation}
w_{A}=\frac{1}{3\gamma^{2}} \,.\label{eq27}
\end{equation}
It is clear from here that the usual expression $P_{A}=w_{A}\rho_{A}=\rho_{A}/3$ cannot be taken \textit{a priori} when the vector field is disformally coupled to other fields.\footnote{The same behaviour occurs when photons experience a disformal transformation \cite{vandeBruck:2013yxa}.} The radiation-like behaviour of the gauge field is, of course, plainly recovered for $\gamma=1$. Hence, it is interesting to investigate some deviations of the above equation and its phenomenological consequences in different stages of the evolution of the universe. We first deal with this issue by determining the dynamical character of the critical points in the phase space and later by performing a numerical analysis to examine the background evolution of the cosmological model.

With all this,  we can define finally the effective equation of state parameter that governs the global dynamics of the universe in terms of the dimensionless variables:
\begin{eqnarray}
w_{eff}\ &=& -1-\frac{1}{3\gamma}\Big[2(\gamma^{2}-1)p^{2}s^{2}+(\gamma^{2}-1)s^{4}+\gamma(\Omega_{m}+\gamma(\gamma^{2}-1)(q^{2}+w^{2}) \nonumber \\
&& -2(2+x^{2}-2y^{2}+\Sigma^{2}))  \Big] \,.\label{eq26}
\end{eqnarray}
%
\section{Cosmological dynamics}
\label{sec:IV}

\subsection{Phase space analysis}\label{subsec:4.1}

To be able to solve the autonomous system in a suitable way and to gain insight into the background cosmology, we have to make some physical and advantageous considerations on the cosmological model. Let us consider the limit $\gamma\to 1$ to start with\footnote{It is worthwhile mentioning that though $\gamma=-1$ is allowed from the dynamical system perspective, this possibility is not  physically suitable according to the definition in Eqn.~(\ref{eq4}).}, which corresponds in fact to the uncoupled limit of the non-Abelian model. As a consequence, $\lambda_{h}$ does not appear explicitly in the fixed points. This translates into a considerable reduction of the phase space dimension. We shall see later that, despite having taken such a convenient choice, we do not lose generality in the description of the main properties of the model since this particular choice corresponds, indeed, to the attractor point of the full dynamical system. Notice, besides, that even though we are preventing an explicit interaction between the scalar field and the Yang-Mills field at this point, they are gravitationally coupled each other, and with the other components, so that differences are expected to arise with respect to the standard quintessence scenario. Another reason why we take this limit is that most of the critical points of cosmological interest are covered in this case. For instance, in the isotropic and the Abelian limits of the model, just a few emergent solutions appear in terms of $\gamma$ (and $\lambda_{h}$), and the other fixed points, which belong to some branches of the uncoupled limit of the non-Abelian model as discussed below, exhibit intriguingly $\gamma=1$. These two limits, still being quite general in this cosmological setting, will be discussed separately to contrast the generality of the results discussed in this part\footnote{It is possible that we may be losing some interesting critical points for the most general case but these are unfortunately inaccessible by the method employed here due to the high dimension of the dynamical system.}. Numerical computations however will enable us to see, in a complementary way to the aforementioned sub-cases, how the disformal coupling operates at the background level in the most general framework. Having clarifying the methodology to be followed, we proceed to analyse the model by accessing some suitable regions of the entire phase space starting from the uncoupled case and going through the coupled cases. As the former case contains most of the critical points of cosmological interest, we will describe it in detail and its derived branches, and focus mainly on new solutions for the other cases.

\subsubsection{The uncoupled case}\label{subsubsec:4.1.1}

We start by describing the uncoupled limit of the non-Abelian model. So, the phase space exhibits 112 fixed points, 48 of them corresponding only to real solutions. Among these, we shall show, however, there exist some general critical manifolds of cosmological interest characterized by a non-vanishing vector field and anisotropic expansion. This means that many branches (submanifolds) are allowed to exist within these general cases, so their main properties will be described in the text. Nevertheless, the stability conditions must be treated separately since the corresponding eigenvalues 
cannot be obtained for the general cases, at least not for all the fixed points. The stability analysis is shown in the appendix \ref{app}. The fixed points found for this case are described as follows:

\begin{enumerate}
\item \textit{Radiation domination}: This fixed point is a \textit{saddle point} describing a radiation era with a scaling behaviour of the scalar and gauge fields with $w_{eff}=w_{\phi}=w_{A}=1/3$. We expect then an extra contribution to the number of relativistic species $N_{eff}$ in the form of non-Abelian dark radiation. This is not a surprising result because the gauge field behaves, in general, as a radiation fluid. Things may be different however when the gauge field is (disformally) coupled to other (scalar) fields while keeping the gamma factor $\gamma\neq1$. The implications of this aspect on the dynamical evolution of the system are discussed in detail at the end of this part and the impact of the gauge field on $N_{eff}$ will be discussed in sec.~\ref{subsec:4.6}. Coming back to the central discussion, we find that four combinations of signs are allowed among the critical points, featuring \textit{isotropic scaling vector-scalar radiation solutions}:
\begin{eqnarray}
&& x= -\frac{2}{\lambda_{V}} \,, \quad y =\pm \frac{\sqrt{2}}{\lambda_{V}} \,, \quad w= \pm\frac{\sqrt{q^{2} + 2 p^{2} s^{2} - 2 s^{4}}}{\sqrt{2}} \,, \nonumber \\
&& \Sigma= 0 \,, \quad \dbtilde{g}= 0 \,, \quad \Omega_{\rm m}= 0 \,, \nonumber \\
&& \Omega_{r}= 1-\frac{3 q^{2}}{2} - 3 p^{2} s^{2} -\frac{6}{\lambda_{V}^{2}} \,.\label{eq28}
\end{eqnarray}
Here the possibility $s=0$ is allowed with $q$ unconstrained (4 solutions). There also exist solutions with $x=y=0$ along with the branches $s=0$ (2 solutions) and $s \neq 0$ (2 solutions), so there are altogether 12 isotropic solutions ($\Sigma{=0}$) taking into account all possible combinations of signs within the different branches. In general, the expressions $\lambda_{V}\neq 0$ and $q^{2} + 2 p^{2} s^{2} > 2 s^{4}$ make up the conditions for the existence of the critical points in phase space. The standard radiation cosmology $\Omega_{r}=1$ is straightforwardly recovered for $x=y=p=s=q=w=0$.

\item \textit{Matter domination}: These fixed points are also classified as \textit{saddle points}, describing a matter dominated era. A set of solutions, called the \textit{anisotropic scaling matter-scalar solutions} with non-vanishing gauge field, are described by the following values:
\begin{eqnarray}
&& x= -\frac{3}{2\lambda_{V}} \,, \quad y =\pm \frac{3}{2\lambda_{V}} \,, \quad w= \pm\frac{1}{4} \sqrt{-3 - 16 s^{4}} \,, \nonumber \\
&& \Sigma= -\frac{1}{4} \,, \quad \dbtilde{g}= 0 \,, \quad \Omega_{r}=0 \,, \quad p=0\,, \quad q=0, \nonumber \\ && \Omega_{m}=\frac{9(-4+\lambda_{V}^{2})}{8\lambda_{V}^{2}} \,.\label{eq29}
\end{eqnarray}
As above, four combinations of signs are allowed among the critical points with the possibility of having, in addition, a vanishing contribution of the scalar field, $x=y=0$ with the same $w$ and $\Sigma$ as above, which implies\footnote{We stress that $\Omega_{m}=9/8$ is attained with no imposition over $\lambda_{V}$ since it arises as an admissible solution of the autonomous system when $x=y=0$.} $\Omega_{m}=9/8$ with a consistent Friedmann constraint due to the gauge field  and dark energy contributions ($\Omega_{\rm DE}=\Sigma^{2}$). Two submanifolds share this general feature because of the possible solutions for $w$ according to Eqn.~(\ref{eq29}). It is interesting to note that the anisotropic shear is always negative 
 which is supported by the non-Abelian nature of the gauge field. Notice however that, for this fixed point, there do not exist physical conditions that guarantee the existence of $w$ for any value of $s$ as can be checked by simple examination. In addition, demanding $\Omega_{m}>0$ leads to a $\lambda_{V}$ out of the range compatible with accelerated expansion (see below). So this point, in its general form, is not of physical interest. 
 
 Another derived solution, which is also admitted in the standard quintessence scenario,  is the \textit{isotropic scaling matter-scalar solution} with vanishing gauge field and with $w_{\phi}=0$:
\begin{eqnarray}
&& x= -\frac{3}{2\lambda_{V}} \,, \quad y =\pm \frac{3}{2\lambda_{V}} \,, \quad w= 0 \,, \quad s= 0 \,, \nonumber \\
&& \Sigma= 0 \,, \quad \dbtilde{g}= 0 \,, \quad \Omega_{r}=0 \,, \quad p=0\,, \quad q=0, \nonumber \\ && \Omega_{m}=1-\frac{9}{2\lambda_{V}^{2}} \,.\label{eq30}
\end{eqnarray}

Another solution gives the standard case $\Omega_{m}=1$  for $x=y=0$. Note that, for all the cases, and unlike the radiation density parameter $\Omega_{r}$, the matter density parameter $\Omega_{m}$ is always independent of the gauge field, though it can contribute to the total energy at this stage. We count ultimately 9 possible critical points during the matter era.

\item \textit{Dark energy domination}: We find 3 general fixed points that can potentially describe the late-time accelerated period of the universe and whose dynamical character must be discussed separately. The first one is characterized by an effective equation of state $w_{eff}=\frac{3(-4+\lambda_{V}^{2})}{12+\lambda_{V}^{2}}$ with different combinations of signs (4 in total) exhibiting novel \textit{anisotropic dark energy solutions}, i.e., with an intriguing presence of the gauge field modulated by the first derivative of the scalar field potential: 
\begin{eqnarray}
&& x = -\frac{6\lambda_{V}}{12+\lambda_{V}^{2}} \,, \quad y =\pm \frac{3\sqrt{2}\sqrt{12-\lambda_{V}^{2}}}{\sqrt{144+24\lambda_{V}^{2}+\lambda_{V}^{4}}} \,, \quad p=0 \,, \quad q=0 \,, \nonumber \\
&& \dbtilde{g}= 0 \,, \quad \Omega_{r}=0 \,, \quad \Omega_{m}=0 \,, \nonumber \\
&& w=\pm \frac{\sqrt{-216-144s^{4}+54\lambda_{V}^{2}-24s^{4}\lambda_{V}^{2}-3\lambda_{V}^{4}-s^{4}\lambda_{V}^{4}}}{\sqrt{144+24\lambda_{V}^{2}+\lambda_{V}^{4}}} \,, \quad \Sigma=\frac{2(-6+\lambda_{V}^{2})}{12+\lambda_{V}^{2}} \,. \nonumber \\
&& \label{eq31} 
\end{eqnarray}
The condition for accelerated expansion reads $-2\sqrt{\frac{3}{5}}<\lambda_{V}<2\sqrt{\frac{3}{5}}$ which is, unfortunately, inconsistent with the condition for the existence of the fixed point itself; in particular, the existence of $w$ demands a more stringent condition than the latter, leading to an unviable alternative. Notice that the branch $s=0$ is allowed but not $w=0$, which does not make things different. Hence, this point can be ruled out as a candidate for addressing the current accelerated expansion in addition to generating large anisotropy $\Sigma\sim\mathcal{O}(0.1)$ (see \cite{Campanelli:2010zx,Saadeh:2016sak,Wang:2017ezt,Akarsu:2019pwn,Amirhashchi:2018nxl}) within the allowed $\lambda_{V}$ values that lead to $w_{eff}<-1/3$.

Another \textit{anisotropic scaling vector solution} is found with $w_{eff}=\frac{-3+\lambda_{V}^{2}}{3+\lambda_{V}^{2}}$ and a gauge field contribution:
\begin{eqnarray}
 && x= -\frac{3\lambda_{V}}{3+\lambda_{V}^{2}} \,, \quad y =\pm \frac{3\sqrt{3}}{\sqrt{9+6\lambda_{V}^{2}+\lambda_{V}^{4}}} \,, \quad s=0 \,, \nonumber \\
 && p=0 \,, \quad w=0 \,, \quad \dbtilde{g}= 0 \,, \nonumber \\
 && \Omega_{r}=0 \,, \quad \Omega_{m}=0 \,, \quad
 q =\pm \frac{3\sqrt{-6+\lambda_{V}^{2}}}{\sqrt{9+6\lambda_{V}^{2}+\lambda_{V}^{4}}} \,, \nonumber \\
 && \Sigma=\frac{6-\lambda_{V}^{2}}{3+\lambda_{V}^{2}} \,.\label{eq32}
\end{eqnarray}
The condition for acceleration reads now $-\sqrt{\frac{3}{2}}<\lambda_{V}<\sqrt{\frac{3}{2}}$ which, once again, is inconsistent with the condition for the existence of the critical point: $\lambda_{V}>\sqrt{6}$ or $\lambda_{V}<-\sqrt{6}$. This point is discarded, as before, as a candidate to address the late-time dynamics.

Finally, we show the standard \textit{quintessence fixed point} which is an attractor solution:
\begin{eqnarray}
&& x= -\frac{\lambda_{V}}{3} \,, \quad y =\pm \frac{1}{3}\sqrt{9-\lambda_{V}^2} \,, \quad s=0 \,, \nonumber \\
&& p=0 \,, \quad q=0 \,, \quad w= 0 \,, \nonumber \\
&& \Sigma=0 \,, \quad \dbtilde{g}= 0 \,, \quad \Omega_{r}=0 \,, \nonumber \\
&& \Omega_{m}=0 \,,\label{eq33}
\end{eqnarray}
with effective equation of state $w_{eff}=-1+\frac{2\lambda_{V}^{2}}{9}$. Thus, the condition for acceleration is simply given by $-\sqrt{3}<\lambda_{V}<\sqrt{3}$, which is consistent with the condition for the existence of the fixed point itself.

\item \textit{Other sets of solutions}:
We also report here, for completeness, other solutions that neither correspond to the standard radiation nor matter nor dark energy dominated epochs. These solutions take place prior to radiation domination when the kinetic term of the scalar field dominates.  They are commonly referred to as \textit{kination}. We report, however, a more general solution modulated by the spatial shear. We call these critical points
\textit{anisotropic kination solutions} with stiff equation of state $w_{eff}=w_{\phi}=1$:
\begin{eqnarray}
&& x= \pm \sqrt{1-\Sigma^{2}} \,, \quad y=0 \,, \quad s=0 \,, \nonumber \\
&& p=0 \,, \quad q=0 \,, \quad w=0 \,, \nonumber \\ && \dbtilde{g}= 0 \,, \quad \Omega_{r}=0 \,, \quad \Omega_{m}=0 \,.\label{eq34}
\end{eqnarray}
Other solutions are given by $\Sigma=\pm1/4$ and $x=\pm\sqrt{15}/4$, $\Sigma=1/2$ and $x=\pm\sqrt{3}/2$, $\Sigma=\pm1$ and $x=0$ (\textit{fully anisotropic domination}), and $\Sigma=0$ and $x=\pm 1$ (\textit{isotropic kination}). 

We finish this part by showing the last critical point which corresponds to the \textit{anisotropic vector domination solution} with $w_{eff}=3$:
\begin{eqnarray}
&& x=0 \,, \quad y=0 \,, \quad s=0 \,, \nonumber \\
&& p=0 \,, \quad q=0 \,, \quad w=\pm\sqrt{-3-s^{4}} \,, \nonumber \\
&& \Sigma=2 \,, \quad \dbtilde{g}= 0 \,, \quad \Omega_{r}=0 \,, \nonumber \\ 
&& \Omega_{m}=0 \,,\label{eq35}
\end{eqnarray}
which is consistent with the phase space conservation, i.e, with the Friedmann constraint.
\end{enumerate}
%

\subsubsection{The isotropic case}\label{subsubsec:4.1.2}

In order to investigate how the disformal coupling operates on the dynamical system, it is very instructive to take, for instance, the isotropic limit of the model $\Sigma=0$. We call the resulting model \textit{disformally coupled isotropic model}. Likewise, to be consistent with the now isotropic Friedmann-Lema\^itre-Robertson-Walker (FLRW) spacetime, we consider the cosmic triad to describe the configuration of the field: that is, mutually orthogonal and of the same norm space-like vector fields with $u(t)=v(t)$ in eqn.~(\ref{eq13}) or, equivalently, $p=s$ and $w=q$ in the jargon of the dynamical system. The Friedmann constraint is consistently modified such that $\Omega_{A}=\gamma(3s^{4}+2q^{2}\gamma^{2})$ as well as all the differential equations are. Before proceeding, we find convenient to make a simple redefinition of the factor $\gamma=1/\beta$ (and later for numerical treatments) to be implemented readily in the autonomous system with the change $\gamma^{\prime}\to\beta^{\prime}$ given by
\begin{equation}
    \beta^{\prime}=\frac{(-1+\beta^{2})(\Pi+x^{2}h)}{x\beta}\,.\label{eq36}
\end{equation}
In doing so, a reduced autonomous system is obtained for which only 26 real solutions are found\footnote{We will consider however the ones with $\beta$ positive by definition.} which are fewer than those found in the uncoupled case. As discussed at the beginning of this section, almost all these solutions have a common fixed point $\beta=1$, belonging thus to some branches of the uncoupled limit of the non-Abelian model once the anisotropy in the background spacetime and in the field itself has been turned off by consistency. So, there is a new version of the \textit{isotropic scaling vector-scalar radiation}:
\begin{eqnarray}
&& x= -\frac{2}{\lambda_{V}} \,, \quad y =\pm \frac{\sqrt{2}}{\lambda_{V}} \,, \quad \dbtilde{g}= 0 \,, \nonumber \\
&&  \beta=1 \,, \quad \Omega_{\rm m}= 0 \,, \quad  \Omega_{r}= 1-\frac{2 q^{2}}{2} - 3 s^{4} -\frac{6}{\lambda_{V}^{2}} \,,\label{eq28a}
\end{eqnarray}
with the possibility of having the branches $q=0$ and $p=s=0$. Other solutions correspond to the isotropic scaling matter-scalar solution eqn.~(\ref{eq30}),  the standard quintessence solution eqn.~(\ref{eq33}) and kination $x=\pm1$, all of them having $\beta=1$. Notice that although the physical features of these fixed points  were already discussed, their associate stability conditions involve now the coupling parameters (see appendix \ref{app1}). 

Interestingly, a new fixed point reveals the presence of the disformal coupling parameters through the gauge field dynamical variables, corresponding to a renewed \textit{isotropic coupling vector-scalar radiation}:
\begin{eqnarray}
&& x= \frac{2}{\lambda_{h}} \,, \quad y =0 \,, \quad q= \pm\sqrt{\frac{-8\beta^{3}+3s^{4}\beta^{2}\lambda_{h}^{2}-3s^{4}\beta^{4}\lambda_{h}^{2}}{2\lambda_{h}^{2}(\beta^{2}-1)}} \,, \nonumber \\
&&  \dbtilde{g}= 0 \,, \quad \Omega_{\rm m}= 0 \,, \quad \Omega_{r}= 1-\frac{4}{\lambda_{h}^{2}}-\frac{8}{\lambda_{h}^{2}(\beta^{2}-1)} \,.\label{eq28b}
\end{eqnarray}
The most intriguing particularity of this solution is the fact that $\lambda_{V}$ is exchanged by the disformal coupling parameter $\lambda_{h}$ in this new version of the radiation dominated period (compare eqn.~(\ref{eq28b}) with eqn.~(\ref{eq28})). Notice also the absence of the scalar potential ($y=0$) in this solution and the early presence of the fields with $w_{\phi}=1$ and $w_{A}=\frac{1}{3\beta^{2}}$ as appropriate. By demanding the existence of this fixed point, the constraints  $\lambda_{h}\neq0$, $0<\beta<1$ (consistent with its own definition), $s<-2^{3/4}\left(\frac{\beta}{3\lambda_{h}^{2}(1-\beta^{2})}\right)^{1/4}\; \lor \; s>2^{3/4}\left(\frac{\beta}{3\lambda_{h}^{2}(1-\beta^{2})}\right)^{1/4}$, are derived. Thus, this is an appealing fixed point supported fully by the coupling of the Yang-Mills field to the scalar field.

\subsubsection{The Abelian case}\label{subsubsec:4.1.3}

It is also interesting to consider the Abelian analog of the model\footnote{The safe way to achieve the Abelian case is to consider solely one component of the gauge field (the one in the $x$-direction) whose residual configuration is still consistent with the axially symmetric Bianchi type-I metric, and not to take intuitively $\dbtilde{g}\to0$. This latter limit, though leads to just one equation of motion as required, lets the last term of eqns~(\ref{eq14}) and (\ref{eq15}) survive, namely the one having two components, which is an unwanted outcome. Thus, once the right limit is taken, the structure of the equations coincides with the one found in the massless case of the disformally coupled Abelian vector model  \cite{Koivisto:2014gia}.} in order to figure out possible emergent critical points when $\beta\neq1$ in a complementary way to the isotropic case. This consideration results in a substantial dimensional reduction of the autonomous system in addition to the simplification of the differential equations. We call this model \textit{disformally coupled Abelian model} which results in an interesting cosmological scenario to be studied. Accordingly, we take the limit $s\to0, q\to0$ and $\dbtilde{g}\to1$ which allows us to perceive some dependence of $\lambda_{h}$ on the fixed points. We do not report here all the found solutions because most of them have counterparts
in the uncoupled case (sec.~\ref{subsubsec:4.1.1}) as discussed early. What is interesting here, however, is that such physical points describing either radiation, matter, or dark energy dominance, belong to the $\beta=1$ branch, but with stability conditions given now in terms of the coupling parameter $\lambda_{h}$ (see appendix \ref{app2}). Thus, the critical points found in the uncoupled case enclose much of the physical properties of the model as can be also verified numerically in sec.~\ref{subsec:4.1.5}. Nevertheless,
we highlight a novel fixed point which corresponds to anisotropic kination with a non-vanishing contribution of the vector field and $\beta\neq1$. In fact, such a critical point depends only on the coupled parameter $\lambda_{h}$; for instance, $\Sigma=\frac{-48+7\lambda_{h}^{2}\pm3\sqrt{-96\lambda_{h}^{2}+9\lambda_{h}^{4}}}{8(6+\lambda_{h}^{2})}$ and $\beta=\frac{-12+2\lambda_{h}^{2}\pm\sqrt{-96\lambda_{h}^{2}+9\lambda_{h}^{4}}}{12+5\lambda_{h}^{2}}$. The full fixed point is quite lengthy to be reported here so we will just discuss its main properties. First, the condition for the existence of this point allows us remarkably to simplify this analysis. It leads to the constraints: $-2\sqrt{3}  <\lambda_{h}<-4\sqrt{\frac{2}{3}}$ and $4\sqrt{\frac{2}{3}}<\lambda_{h}<2\sqrt{3}$. For the sake of comparison, let us focus on the latter range. The lower limit of it provides a vector contribution of $\Omega_{A}\approx 0.37$ and the upper limit provides $\Omega_{A}\sim 10^{-6}$. In the latter case, the fixed point can be well approximated to  $x\approx\frac{\sqrt3}{2}, \Sigma\approx\frac{1}{2}, \beta\approx\frac{1}{\sqrt{3}}$. We notice, however, that the lower limit of $\lambda_{h}$ reduces the shear up to a factor 2, it still being large compared with the present value. Notice however that, during this period, larger levels of anisotropy than the current one are, in principle, allowed. We can conclude that the upper limit case encloses the main properties of the anisotropic kination solution found in the non-Abelian case. One can also ask whether accelerated expansion can be driven by this point given the suggestive form of the effective equation of sate: $w_{eff}=-1+\frac{57\lambda_{h}^{4}-36\chi+3\lambda_{h}^{2}(-156+7\chi)}{6(6+\lambda_{h}^{2})(-12+2\lambda_{h}^{2}+\chi)}$, where $\chi=\sqrt{-96\lambda_{h}^{2}+9\lambda_{h}^{4}}$. It requires however $\lambda_{h}=0$ which cannot be consistent with the condition for the existence of the critical point itself. 

We also find a reduced version of the anisotropic kination solution of the uncoupled case, supported by the disformal coupling $\lambda_{h}$ instead of the potential parameter $\lambda_{V}$. It reads
\begin{eqnarray}
&& x= \frac{3}{\lambda_{h}} \,, \quad y =0 \,, \quad w= 0 \,, \quad p= 0 \,, \nonumber \\
&& \Sigma= \pm \sqrt{1-\frac{9}{\lambda_{h}^{2}}} \,, \quad \Omega_{\rm m}= 0 \,.\label{eqabel0}
\end{eqnarray}
The existence of this point is plainly guaranteed for $\lambda_{h}>3\; \lor \; \lambda_{h}<-3$. Hence, the level of anisotropy in the expansion can be as small as one wishes in absolute value, the former being either positive or negative. Other solutions with negative anisotropy are the fully anisotropic solution $\Sigma=-1$ and the kination solution
\begin{eqnarray}
&& x= -\frac{\sqrt{15}}{4} \,, \quad y=0 \,, \quad w= 0 \,, \nonumber \\
&& \Sigma= -\frac{1}{4} \,, \quad \beta= 1 \,, \quad \Omega_{\rm m}= 0 \,,\label{eqabel01}
\end{eqnarray}
in addition to the already positive anisotropy kinetic solutions discussed in the uncoupled limit. Here the branch $p=0$ is also allowed. On the other hand, the general radiation point does not admit the gauge field contribution but does admit the contribution of the scalar field:
\begin{eqnarray}
&& x= -\frac{2}{\lambda_{V}} \,, \quad y =\pm \frac{\sqrt{2}}{\lambda_{V}} \,, \quad w= 0 \,, \nonumber \\
&& \Sigma= 0 \,, \quad \beta= 1 \,, \quad \Omega_{\rm m}= 0 \,, \nonumber \\
&& \Omega_{r}= 1 -\frac{6}{\lambda_{V}^{2}} \,.\label{eqabel1}
\end{eqnarray}
Here it is also possible to have the branch $x=y=0$ that leads to the standard radiation scenario. As to the matter solution, it admits gauge field contributions in addition to anisotropy. We can call this fixed point, similar to the general case, \textit{anisotropic scaling matter-scalar solution} with non-vanishing gauge field. It reads
\begin{eqnarray}
&& x= -\frac{3}{2\lambda_{V}} \,, \quad y =\pm \frac{3}{2\lambda_{V}} \,, \quad w= \pm\frac{\sqrt{3}}{4} \,, \nonumber \\
&& \Sigma= -\frac{1}{4} \,, \quad p=\pm\sqrt{\frac{2}{3}} \,, \quad \Omega_{r}=0 \,, \nonumber \\ && \beta=-1  \,,   \quad  \Omega_{m}=\frac{9(-4+\lambda_{V}^{2})}{8\lambda_{V}^{2}} \,.\label{eqabel2}
\end{eqnarray}
The branch $w=p=0$ is also allowed. As in the non-Abelian case, demanding $\Omega_{\rm m}>0$ leads to a $\lambda_{V}$ out of the allowed range to perform accelerated expansion. We are left then with the standard matter-dominated solution. In addition, notice that $\beta$ negative is not allowed (see footnote 4). The reason why we report this unphysical fixed point here is to evidence the nonexistence of the Abelian version of the matter dominated solution given by eqn.~(\ref{eq29}), that is, the one with negative anisotropy in the expansion. 

Another solution embedded in the uncoupled case is the scaling matter-scalar solution, which has a similar form as eqn.~(\ref{eq30}):
\begin{eqnarray}
&& x= -\frac{3}{2\lambda_{V}} \,, \quad y =\pm \frac{3}{2\lambda_{V}} \,, \quad w= 0 \,, \nonumber \\
&& \Sigma= 0 \,, \quad p= 0 \,, \quad \Omega_{r}=0 \,, \nonumber \\ && \beta=1 \,, \quad \Omega_{m}=1-\frac{9}{2\lambda_{V}^{2}} \,.\label{eq30b}
\end{eqnarray}
Nevertheless, its dynamical character may change depending on the associated eigenvalues from a saddle to an attractor point (see appendix \ref{app2}). However, it always corresponds to matter domination $w_{eff}=0$, such that it is not possible to generate accelerated expansion.

Lastly, we have found the \textit{anisotropic Abelian vector-scalar solution}
\begin{eqnarray}
&& x = -\frac{6\lambda_{V}}{12+\lambda_{V}^{2}} \,, \quad y =\pm \frac{3\sqrt{24-2\lambda_{V}^{2}}}{12+\lambda_{V}^{2}} \,, \quad p=\pm\frac{(12+\lambda_{V}^{2})\sqrt{72-18\lambda_{V}^{2}+\lambda_{V}^{4}}}{\sqrt{6}(-6+\lambda_{V}^{2})(12+\lambda_{V}^{2})} \,, \nonumber \\
&& \beta= 1 \,, \quad \Omega_{r}=0 \,, \quad \Omega_{m}=0 \,, \nonumber \\
&& w=\pm \frac{\sqrt{3}\sqrt{72-18\lambda_{V}^{2}+\lambda_{V}^{4}}}{12+\lambda_{V}^{2}} \,, \quad \Sigma=\frac{2(-6+\lambda_{V}^{2})}{12+\lambda_{V}^{2}} \,, \nonumber \\
&& \label{eqabel3} 
\end{eqnarray}
with effective equation of state $w_{\rm eff}=-1+\frac{4\lambda_{V}^{2}}{12+\lambda_{V}^{2}}$ from which the condition of accelerated expansion $-2\sqrt{\frac{3}{5}}<\lambda_{V}<2\sqrt{\frac{3}{5}}$ is derived. This condition is compatible with the one demanded to guarantee the existence of the critical point  $-\sqrt{6}<\lambda_{V}<\sqrt{6}$. Notice, however, that it was not possible to reconcile both constraints in the non-Abelian case. Unfortunately, the degree of anisotropy in the expansion $\Sigma$, within the allowed range of values for $\lambda_{V}$, is two orders of magnitude larger (that is $\Sigma\sim\mathcal{O}(0.1)$) than the current value inferred by observations. Accordingly, the trajectories leaving  possible anisotropic radiation points will approach to an isotropic matter-dominated solution, though it is possible to take adequate initial conditions to prevent the rapid dilution of the anisotropy as can be seen in the numerical analysis. This statement is valid from the general case to the particular cases studied here.

On the other hand, a dark energy dominated solution corresponds to the standard quintessence scenario with $\beta=1$ over again but, contrary to the non-Abelian case, the  stability conditions do place constraints over $\lambda_{h}$ even though the full critical point does not exhibit explicit dependence on it. This is shown in appendix \ref{app2}.

In summary, we have confirmed that most of the resulting fixed points in the isotropic and Abelian cases correspond to a submanifold of the anisotropic non-Abelian uncoupled case once the appropriate limits are taken. There do exist, however, for both cases some genuine fixed points that can be interesting for early cosmological scenarios. Hence, this detailed analysis supports the fact that the  phase space studied in the uncoupled case ($\beta=\pm1$) together with the other two cases encompass most of the global properties of the model even though we are not fully covering the phase space due to its complexity. 

\subsection{Phase space trajectories}\label{subsec:4.1.5}

We have classified so far the fixed points found according to their cosmological character in three different groups or cases: the uncoupled, the isotropic, and the Abelian limits of the model, covering thus most of the entire phase space. Indeed, most of the fixed points were found in the uncoupled case with some branches belonging to the isotropic and the Abelian cases. It means that these two latter cases are roughly self-contained cases of the uncoupled limit except for a few new emergent solutions where $\beta\neq1$. Thus, we have shown that $\beta=1$ is a fixed point in most of the solutions found and corresponds ultimately to the common attractor point for the three cases studied. An interesting inquiry that arises is then whether this situation really corresponds to an attractor solution of the full dynamical system such that the above choice can be fully justified within the entire phase space. To answer this question, we have to describe in a general way the evolution in phase space of the autonomous systems and to look at specifically how $\beta$ evolves. This is done by numerically evolving parametric trajectories in phase space for different initial conditions. 

\begin{figure*}
\includegraphics[width=0.45\hsize,clip]{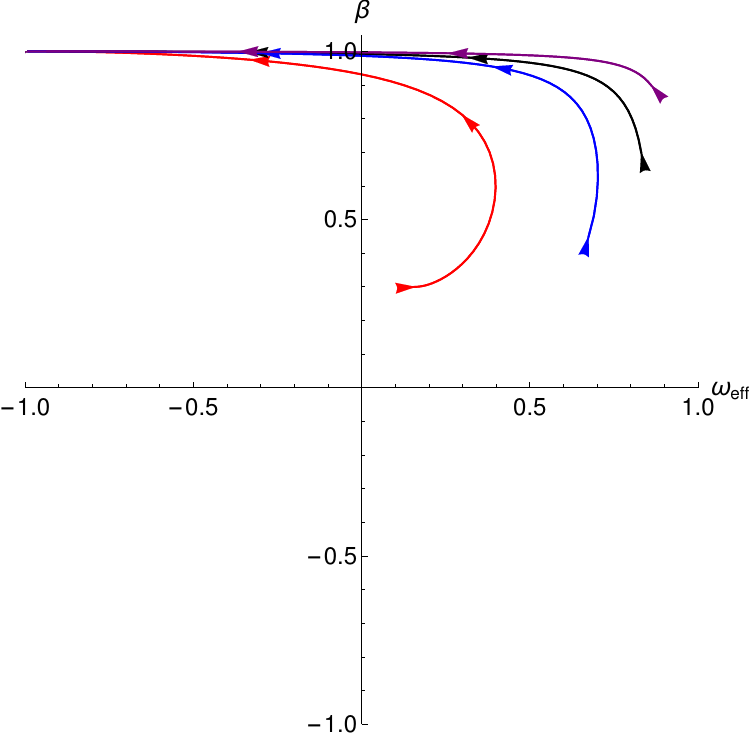}
\includegraphics[width=0.55\hsize,clip]{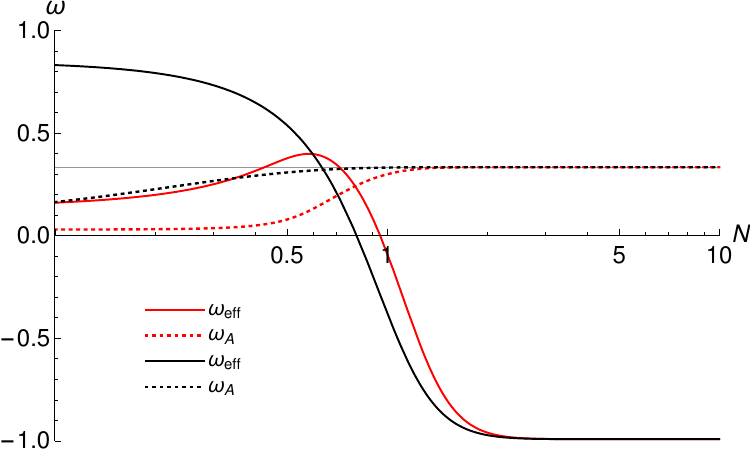}
\caption{Left panel: parameter space $\beta-w_{eff}$ for four different initial conditions. We have chosen $\lambda_{V}=0.2, \lambda_{h}=0.7, y_{i}=p_{i}=s_{i}=w_{i}=q_{i}=\Sigma_{i}=\Omega_{m,i}=0.1$, and $\Omega_{r}=0$ for all the curves
and $x_{i}= 0.923164, \beta_{i}=0.9$ for the purple curve, $x_{i}= 0.906235, \beta_{i}=0.7$ for the black curve, 
$x_{i}=0.812314, \beta_{i}=0.45$ for the blue curve, and $x_{i}= 0.371832, \beta_{i}=0.3$ for the red curve. Right panel: evolution of $w_{eff}$ (solid curves) and $w_{A}$ (dotted curves) versus the number of e-folds $N$;  the horizontal grey line corresponds to $w_{A}=1/3$; we have shown, for illustrative purposes, just two particular cases given by the initial conditions associated with the red and black curves, respectively, as in the left panel.} \label{fig:0}
\end{figure*}

The dynamical evolution of the system is displayed in the left panel of Fig.~\ref{fig:0} for different sets of initial conditions. In particular, we show the parameter space $\beta-w_{eff}$ following  different trajectories in phase space which evidence that $\beta=1$ corresponds to an attractor solution. The black line corresponds to an initial value $\beta_{i}=0.7$, it being somehow far enough from $\beta=1$ to allow a brief evolution of the system before reaching the attractor stable solution $(w_{eff},\beta)=(-1,1)$.  In contrast, a little further away from the attractor point, the red line denotes a set of initial conditions with $\beta_{i}=0.3$.
Accordingly, the simple fact of assigning values to $\beta_{i}$ close or far from $1$ leads to some interesting implications for the role of the vector field during the cosmological evolution. This is observed from the evolution of the equation of state $w_{A}$ in the right panel of Fig.~\ref{fig:0}.  We can see in this panel that the red-dotted solution experiences a longer transition during which $w_{A}\approx0$, i.e., the gauge field acts as a matter fluid before reaching the radiation-type nature characterized by $w_{A}=1/3$ shortly after 1 \textit{e-fold}. We have a very different situation if the trajectory rapidly reach the attractor solution (black dotted line) in which case no matter-type behaviour is exhibited. This depends, of course, on how close or far $\beta_{i}^2$ is from unity since $w_{A}=\beta^2/3$. We shall see, however, that for very small initial values of $w$ and $q$, the condition $\beta_{i}^2$ close to $1$ is not a sufficient condition for reaching rapidly the radiation behaviour but it depends generally on the other initial conditions. This latter aspect shall be investigated numerically in more detail in the next part. 

Going back to the main discussion, we have shown that $\beta=1$ is undoubtedly a future attractor point of the full dynamical system, so the generality of the previous results as well as the main features of the critical points described above are granted. One can think, however, that the choice $\beta=1$ can be justified only for the late-time dynamics and not, for instance, during the radiation and matter eras. We have shown however that there also exist solutions where $\beta=1$ is a fixed point in most of the solutions describing those periods. Yet, we will tackle this issue in the physical space, in the next part, by solving numerically the cosmological background evolution of the model.
%

\subsection{Cosmological background evolution: numerical results}\label{subsec:4.3}
We shall consider the general case ($\beta\neq 1$) to analyze the behaviour of the background cosmology of the model
in order to check that
its main features 
do indeed correspond to those already described in sec. \ref{subsec:4.1}.
We 
will do it
by performing some numerical computations for different values of the model parameters and initial conditions. As a result,
we have observed that there are no significant changes on the global dynamics when varying $\lambda_{h}$ for different sets of initial conditions.
On the other hand, the effects of $\lambda_{V}$ have a mostly visible impact, as in the standard quintessence case, on the late-time dynamics, quantified by the effective equation of the state $w_{eff}=-1+\frac{2\lambda_{V}^{2}}{9}$ in accordance with the attractor fixed point. Hence, $\lambda_{V}$ will be kept fixed for all the computations but with $\lambda_{V}\neq0$ so that we can focus on the other parameters that are associated to the gauge field to see their impact on the overall evolution. On the other hand, as we have an extra component in the total budget of the universe given by the vector field and quantified by $\Omega_{A}$, it will be crucial to determine its contribution during different stages in the evolution of the universe in comparison to the $\phi$CDM reference model. 
\begin{figure}[!htb]
\includegraphics[width=0.50\hsize,clip]{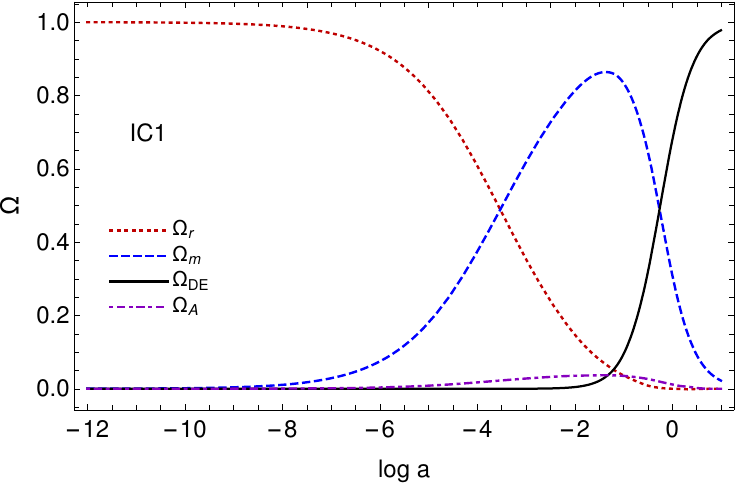}
\includegraphics[width=0.50\hsize,clip]{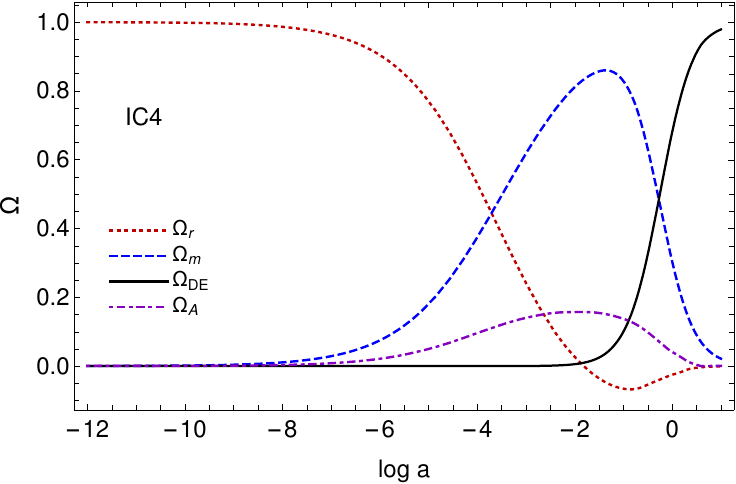}
\includegraphics[width=0.50\hsize,clip]{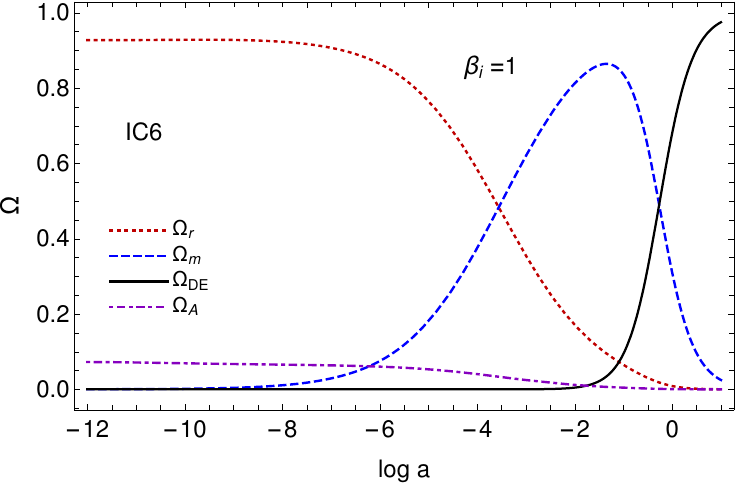}
\includegraphics[width=0.50\hsize,clip]{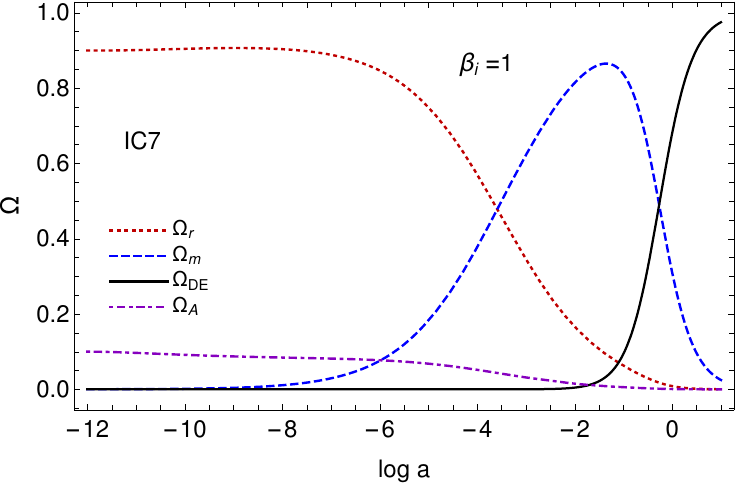}
\caption{Evolution of the density parameters $\Omega$ versus the number of e-folds $N \equiv \log{a} = \alpha$ for four different sets of initial conditions as described in the main text. Top panel: these solutions correspond to the $\beta_{i}\neq1$ case where $\Omega_{A}$ can contribute lately. Bottom panel: these solutions, on the contrary, show the $\beta_{i}=1$ case where $\Omega_{A}$ can contribute mainly during the radiation dominated epoch. All the initial conditions presented in the main text have been chosen to match the present values $\Omega_{DE}^{(0)}\approx 0.7, \Omega_{m}^{(0)}\approx 0.3$ \cite{Abbott:2017wau}.} \label{fig:1}
\end{figure}
After exploration of the initial conditions, we will notice that the global behaviour of the density parameters is quite sensible to the initial values of the vector field components $p$ and $s$ and their velocities $w$ and $q$.  This allows the model to exhibit new features whose imprints on cosmological observables at different epochs can be tested with different cosmological surveys. Interestingly, we will see that the initial value of the factor $\beta$, which encodes information about the disformal coupling, can change drastically the physical character of the equation of state parameter of the vector field $w_{A}=\beta^{2}/3$, the radiation-like feature being recovered for $\beta_{i}=1$ as we anticipated in the phase space analysis. Indeed, this case corresponds to a critical point of the system but, as we discussed, an interesting transient situation can be attained if we start with an initial value for $\beta$ different from 1 depending strongly on the initial conditions for the other variables as well, leading clearly to a matter-like equation of state.

To see how these features can be distinguished each other, let us first show the $\beta_{i}\neq1$ case for different sets of initial conditions (IC). In particular, we focus on changes in the initial values for $q$ and $w$ to make this analysis more manageable. We call these ICi solutions, with i running from 1 to 5, for a better illustration of how they work on the global cosmological dynamics. We set different sets of initial conditions at $N=-12$ well within the deep radiation-dominated era. So, we consider the fiducial values: $\lambda_{V}=0.5, \lambda_{h}=1.2, x_{i}=2\times10^{-12}, y_{i}=3\times10^{-10}, p_{i}=10^{-2}, s_{i}=10^{-2}, \Sigma_{i}=10^{-3},
\dbtilde{g}_{i}=10^{-6}$ in addition to IC1=($w_{i}=6\times10^{-4}, q_{i}=10^{-3}, \beta_{i}=0.5, \Omega_{r,i}=0.999789$), IC2=($w_{i}=10^{-4}, q_{i}=7\times10^{-3}, \beta_{i}=0.9, \Omega_{r,i}=0.999789$), IC3=($w_{i}=4\times10^{-5}, q_{i}=10^{-2}, \beta_{i}=0.3, \Omega_{r,i}=0.996095$), IC4=($ w_{i}=4\times10^{-5}, q_{i}=2\times10^{-2}, \beta_{i}=0.99, \Omega_{r,i}=0.999387$), and IC5=($ w_{i}=4\times10^{-5}, q_{i}=2\times10^{-2}, \beta_{i}=0.3, \Omega_{r,i}=0.984984$). They all lead to a consistent cosmological evolution with the expected radiation, matter and dark energy dominated periods. Notice that we take $\lambda_{h}$ fixed when solving numerically because small differences in it lead to very small percent differences in the numerical evolution of all the studied quantities. So, the effect of changing the disformal parameter on the cosmological evolution is not appreciable in any plot.
\begin{figure*}
\includegraphics[width=0.50\hsize,clip]{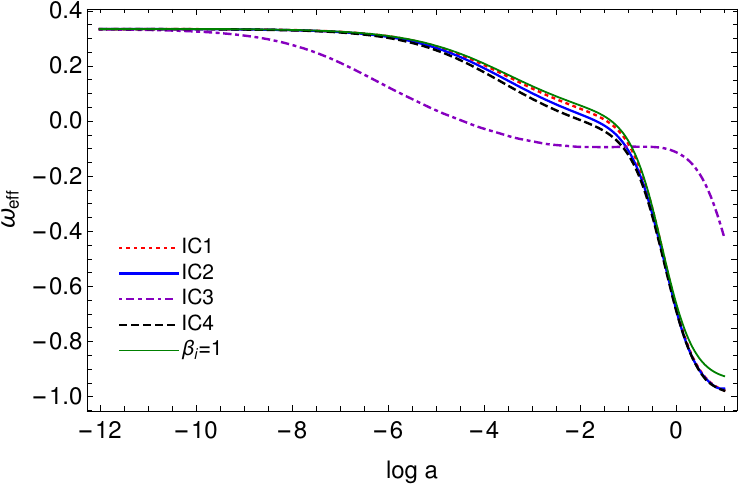}
\includegraphics[width=0.50\hsize,clip]{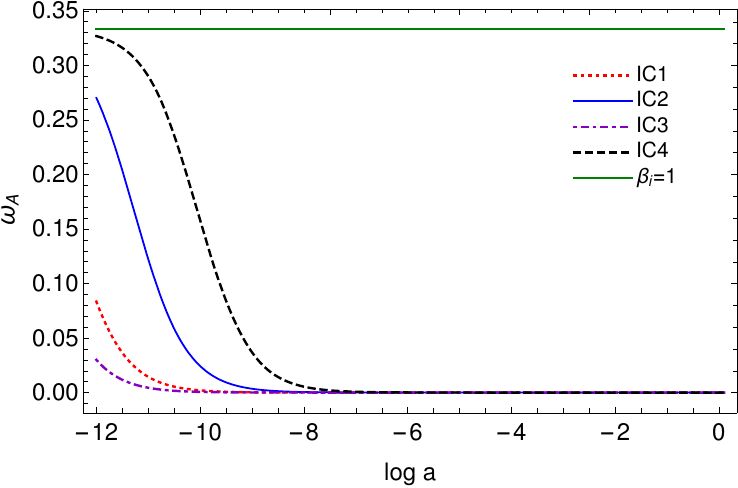}
\caption{Left panel: evolution of $w_{eff}$ versus the number of e-folds $N \equiv \log{a} = \alpha$ for different initial conditions characterized mainly by $\beta_{i}\neq1$ (ICi) and $\beta_{i}=1$ (IC6 solution denoted by the green solid line) cases as described in the main text. Right panel: evolution of $w_{A}$ versus $N$ for the same cases shown in the left panel.} \label{fig:2}
\end{figure*}

Thus, we plot the density parameters of all the involved components for the $\beta_{i}\neq1$ case in the top panels of Fig.~\ref{fig:1}. We prefer to show here only the two particular cases IC1 and IC4 since they expose the general features of this case whereas the other cases are left for later illustration. As can be inferred, the effects of the coupling start to be appreciable around $N\approx -6$ (at the nucleosynthesis epoch), much before matter-radiation equality, with very subtle differences on the redshift equality, in comparison to the standard $\Lambda$CDM model, that can be crucial in determining a consistent cosmic history according to observations. A more compelling feature, as a consequence of the disformal coupling however, occurs during the matter-dominated era: there, even though the vector field density parameter is quite subdominant, its contribution is clearly significant 
to impact the formation of large-scale structures, as can be highlighted for the numerical computation IC4 (top right panel of the same figure). The enhancement of $\Omega_{A}$ is produced however at the expense of $\Omega_{r}$, leading the latter to acquire a negative value which might not correspond to a physical situation. 
To better understand why $\Omega_{r}$ may be negative, we have first to notice that initial conditions are taken in such a way that the energy density parameters match the present values and the contribution of the gauge field to the total energy density is significant. We are able to do this because there are simply many variables to set the entire initial conditions that we can give as much energy density to the gauge field as possible. The increasing of energy is possible because $\beta$ is very small during the transient solution, that is much before it approaches the attractor point, and the energy density associated to the gauge field scales as $\beta^{-3}$ when it becomes a maximum. As a consequence,  there must be a negative energy density to compensate the exaggeratedly extra (positive) contribution of the gauge field via the Friedmann constraint which sets the gravitational interaction between all components. Nevertheless, $\Omega_{m}$ cannot decrease drastically since matter-radiation equality and $\Omega_{m}^{(0)}\approx0.3$ must be guaranteed. On the other hand, $\Omega_{DE}$ must be an increasing function to match its current value. In this regard, we are somehow forcing the system to have, as the only alternative, $\Omega_{r}$ negative to satisfy the aforementioned constrains. This is of course a harmful effect we want to prevent even though initial conditions match the present values. It does not mean, however, that $\beta_{i}$ can be constrained from this analysis \textit{a priori} but, rather, it exposes the complexity of choosing different and suitable initial conditions still being consistent with the cosmological evolution. Hence, the safest way to tackle this problem 
is to
use the observational data due to the large number of involved variables in the numerical analysis. This is a mandatory challenge we have to deal with in a future investigation.

We perform now numerical computations for the $\beta_{i}=1$ case taking the same fiducial values as above with IC6=($w_{i}=10^{-1}, q_{i}=2.5\times10^{-1}, \beta_{i}=1, \Omega_{r,i}=0.927297$) and IC7=($w_{i}=10^{-1}, q_{i}=3\times10^{-1}, \beta_{i}=1, \Omega_{r,i}=0.899797$). The results are displayed in the bottom panel of Fig.~\ref{fig:1}. It is distinctive now the early contribution of the vector field in the form of dark radiation but still contributing during the matter domination until the scalar field starts growing. We quantify the contribution of the vector field to the total energy density during the deep radiation era to be around $7.3\%$ and $10\%$ for IC6 and IC7 numerical solutions, respectively. Accordingly, these extra degrees of freedom may contribute to the total number of relativistic species.  In addition, they may affect some cosmic microwave background (CMB) features 
which become further cosmological consequences of the model 
we must keep in mind. Hence, this physical scenario has to face several cosmological constraints in the search of its phenomenological viability.

Likewise, the effective equation of state also shows the general behaviour of the model following the expected sequence, radiation-matter-dark energy dominated epochs. For this case, slight differences are shown between the $\beta_{i}\neq1$ and $\beta_{i}=1$ (IC6) cases, particularly during the onset of matter domination\footnote{Notice that the initial conditions we have chosen yield actually a non-fully matter dominated epoch, i.e., $\Omega_{m}<1$, aiming at highlighting the role of the gauge field during the cosmological evolution.}  and the present value of $w_{DE}$ as is appreciated in the left panel of Fig.~\ref{fig:2}. It implies that $w_{DE}$ is not only established by the parameter $\lambda_{V}$, as in the $\phi$CDM model, but a degeneracy is manifested as a consequence of taking different and suitable initial conditions to describe the cosmic history.\footnote{In the future, however, $w_{eff}=-1+\frac{2\lambda_{V}^{2}}{9}$ according to the attractor solution described in sec.~\ref{subsec:4.1}.} On the other hand, the IC3 computation ($\beta_{i}=0.3$), which corresponds to a stronger coupling than the other displayed cases, along with a value for $q$ which is an order of magnitude smaller than those of the other solutions, is shown here for illustration purposes only since it does not lead to a consistent cosmological dynamics as it happens with IC4\footnote{Notice, once again, that though small values of $\beta_{i}$ lead potentially to an unphysical situation, we can fix it with an appropriate choice of the other (e.g, $w_{i}$ or $q_{i}$) initial conditions.}. As an unequivocal consequence of taking such a small value for $\beta_{i}$, $w_A=\beta^{2}/3$ approaches zero during almost all the cosmological evolution, as can be seen in the right panel of Fig.~\ref{fig:2}, but this fact is of no concern, in general, provided that it leads to well-defined energy densities for the remaining components. 
Thus, $w_A$ may transit from $0$ to $1/3$ provided that $\beta_{i}\neq1$ (and small) as can be glimpsed in Fig.~\ref{fig:1}. Notice however that the full transitions to the attractor point $w_{A}=1/3$ ($\beta=1$) for the $\beta_{i}\ne1$ cases are not seen in Fig.~\ref{fig:2} because these happen after the present time when integration has already stopped. This qualitative feature is strongly sensitive to the initial conditions that must be chosen such that they match the current values of the density parameters. In physical terms, the gauge field behaves as dark matter for most of the evolution of the universe provided that $\beta_{i}\neq1$, but in the future it will recover the standard form $w_{A}=1/3$ since, as we argued, $\beta=1$ is an attractor point of the system. We have not observed in the numerical analysis that the transition may occur before the present time for different sets of initial conditions. In contrast, the vector field exhibits the standard radiation-like behaviour, as we have explored but not shown here, for a wide range of initial conditions with $\beta_{i}=1$, i.e., $w_{A}=1/3$ (see the green solid line corresponding to the numerical solution IC6), during all the cosmological evolution by virtue of the attractor-type solution. The property of behaving like a non-relativistic fluid (say before the time of matter-radiation equality) is analogue to the one provided by successful vector dark matter models once the field acquires mass by some available mechanism (see e.g. \cite{Dror:2018pdh}). But on the other hand, our scenario portraits a massless vector field whose coupling $\beta$ acts somehow as an effective mass term before reaching the attractor point $\beta=1$. This is an interesting prediction of the model that deserves a comprehensive investigation at perturbative level during the formation and evolution of large-scale structures (LSS).

\begin{figure*}
\includegraphics[width=0.50\hsize,clip]{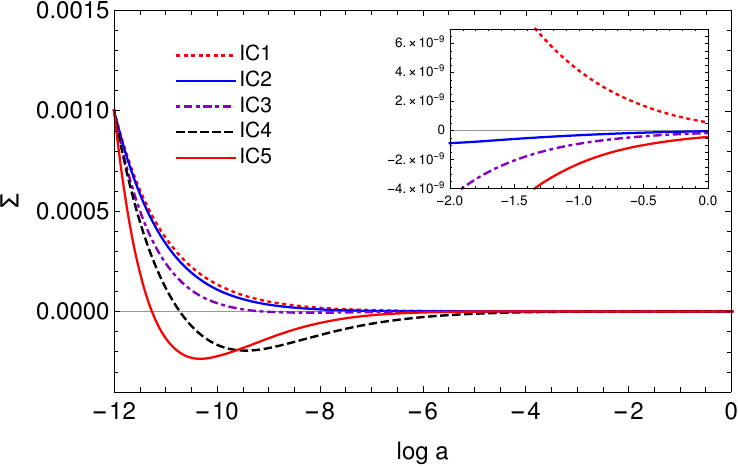}
\includegraphics[width=0.50\hsize,clip]{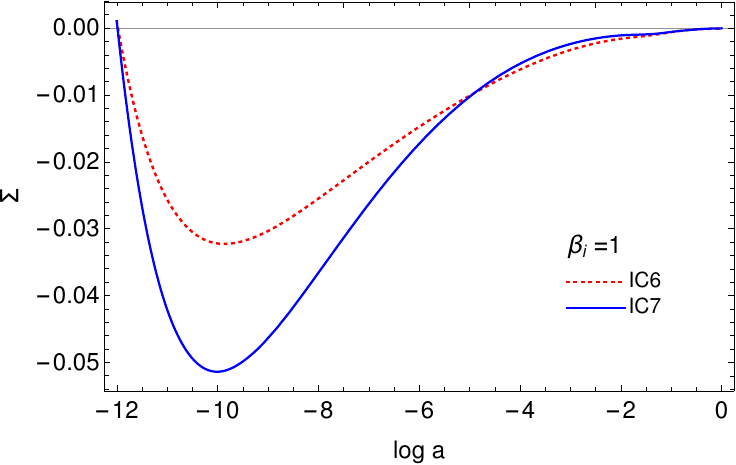}
\caption{Evolution of the spatial shear versus the number of e-folds $N \equiv \log{a} = \alpha$ for different initial conditions. The left panel corresponds to numerical solutions with $\beta_{i}\neq1$ while the right panel shows numerical solutions for the $\beta_{i}=1$ case.
} \label{fig:3}
\end{figure*}

Finally, we evaluate the degree of anisotropy in the expansion by plotting the spatial shear for different initial conditions as before. The results for $\beta_{i}\neq1$ are shown in the left panel of Fig.~\ref{fig:3}. We can see that the shear rapidly decreases going to very small but non-zero values of $\mathcal{O}(\pm10^{-9})$ at the present time with a maximum value at the radiation era which depends, of course, on the initial conditions. What is interesting here is the possibility of having both positive and negative values of $\Sigma$ during all the cosmological evolution which is a distinctive characteristic supported by non-Abelian vector fields in contrast to the Abelian case (see sec.~\ref{subsubsec:4.1.3}). For instance, the dotted-red curve (IC1 solution) provides at the present time $\Sigma_{0}>0$; on the contrary, the other solutions give  $\Sigma_{0}<0$. On the other hand, the $\beta=1$ case accounts for $\Sigma_{0}<0$ for most of the chosen initial conditions (see right panel of Fig.~\ref{fig:3}). We can conclude, therefore, roughly speaking, that the sign of $\Sigma$ is determined somehow by how close (negative) or far (positive) $\beta_{i}$ is from 1. We finish this part by estimating the order of magnitude of $\Sigma_{0}$ for the $\beta_{i}=1$ case to be of $\mathcal{O}(-10^{-5})$ larger that the $\beta_{i}\neq1$ case, which means that the mechanism to produce anisotropy is more efficient in the former case. As advertised, this analysis is performed by keeping $p$ and $s$ fixed. However, we notice that increasing $p_{i}$ by a factor 5, for instance, implies that the shear becomes $\mathcal{O}(-10^{-4})$. This shows once again the difficulty of unveiling the main features of the model due to the large number of involved variables and supports the procedure we have carried out in the numerical analysis. Despite all the discussed possibilities to find a non-vanishing shear in the late universe, the results are well below the bound set by the magnitude-redshift data of type Ia supernovae $|\Sigma_{0}| \leq\mathcal{O}(10^{-3})$ \cite{Campanelli:2010zx,Saadeh:2016sak,Wang:2017ezt,Akarsu:2019pwn,Amirhashchi:2018nxl}. This bound can be, however, much lower when the baryonic acoustic oscillations (BAO) and CMB data are included, but it has to be done in a model-dependent way as was recently inferred within the $\Lambda$CDM concordance model \cite{Akarsu:2019pwn}. 

Another important constraint on the anisotropic expansion comes from the power spectrum of temperature fluctuations
in the CMB data when free streaming photons are propagating, since the recombination epoch up to the present time, for different expansion factors \cite{Koivisto:2008ig,Appleby:2009za}. The value observed by the Planck satellite is $\Delta T_{\rm Planck}\approx 14\mu K$ which is lower than the one predicted by the $\Lambda$CDM model \cite{Aghanim:2018eyx}. This discrepancy is referred to as the quadrupole temperature problem and opens up the possibility of additional contributions to $\Delta T$. For instance, the anisotropic expansion can contribute to the CMB  temperature fluctuation as $\Delta T_{\Sigma}\propto |\sigma(t_{0})-\sigma(t_{\rm dec})|\sim\mathcal{O}(10^{-4})$ where $t_{\rm dec}$ and $t_{0}$ correspond to the cosmic time at decoupling and today, respectively \cite{Appleby:2009za}. Hence, it will be challenging to include all the aforementioned constraints for the anisotropic expansion in a model dependent way and at different redshifts to check the cosmological viability of the model.

In summary, we have assessed the role of the Yang-Mills field during the cosmological evolution and its contribution to the total energy density during different epochs. The global dynamics shows interesting differences with respect to the uncoupled $\phi$CDM model that may help better describe the observational data. It remains to perform a full exploration of the initial conditions space without assuming fiducial values but this can be addressed in a more suitable way by including background observational data such as supernovae data and measurements of the CMB. For instance, the present expansion rate $H_{0}$ inferred by measurements of the temperature fluctuations, polarization, and lensing in
the CMB radiation, depends sensitively on the expansion history before and after recombination. This is then a crucial caveat that any cosmological model, including the one proposed here, must overcome to ensure its cosmological viability. The early presence of the gauge field may also help alleviate the Hubble tension \cite{Freedman:2017yms}, leaving the late-time dynamics untouchable or with little impact on local measurements of $H_{0}$.
%

\subsection{Phenomenological couplings}\label{subsec:4.5}
Another possibility to affect the dynamical behaviour of the model is to consider phenomenological couplings, then making more explicit the functional form of both the conformal $C(\phi)$ and the disformal $D(\phi)$ couplings as was done in \cite{vandeBruck:2015tna}. The main idea behind this is to allow $\gamma$ 
to evolve differently such that it may impact the global dynamics.  

If we consider for instance
\begin{equation}
    C(\phi)=C_{0} e^{c\phi},\; D(\phi)=D_{0} e^{d\phi} \,,\label{eq37}
\end{equation}
with $C_0,D_0,c$, and $d$ being constants,
the $\gamma$ factor takes then the form $\gamma=\left(1-\theta x^{2}\right)^{-1/2}$ with $\theta \equiv 6 M_p{^2} H^{2} \frac{D}{C}$ as a new dimensionless variable of the autonomous system whose evolution equation is
\begin{equation}
    \theta^{\prime}= 2\theta(x\lambda_{h} -\epsilon_{H}) \,.\label{eq38}
\end{equation}
%
The new form of the autonomous system, however, prevents us from finding analytical critical points even in the Abelian case, so we will describe here some qualitative results based mainly on numerical computations. A promising change we notice immediately is the new dependence  of the vector field equation of state on the kinetic energy of the scalar field $x$ as $w_{A}=\frac{1}{3}(1-\theta x^{2})$. Nevertheless, after numerical exploration, we find that $\theta=0$ is an attractor solution of the dynamical system as we suspected. It implies consequently that $w_{A}=1/3$ during all the cosmological evolution despite choosing initial conditions for $x$ large enough to hopefully allow $\theta x^2$ to evolve. 
Furthermore, we realize that, despite this new choice, it is possible to write linearly $\lambda_{h}$ as a sum of $\lambda_{D}\propto\frac{D_{\phi}}{D}$ and $\lambda_{C}\propto\frac{C_{\phi}}{C}$, leading therefore to a single effect of the coupling functions. Hence, this case is analogue to the already investigated $\beta_{i}=1$ case, so it closely reproduces the well-known cosmological properties found before. This is the reason why we do not show additional numerical results. 
The only important difference we find is that the parameter $\lambda_{h}$ affects notoriously the behaviour of $\theta$ before the latter goes to zero, it being unimportant anyway in order to notice significant changes in the global dynamics except for the kination era where $x$ dominates. The impact of the vector field coupling during such a period may relax the current bound $\Omega_{DE}< 0.045$ set by big bang nucleosynthesis (BBN) measurements \cite{Battye:2009ze}, so it is an interesting issue that may be relevant.

\subsection{Non-Abelian dark radiation and $N_{eff}$}\label{subsec:4.6}

\begin{figure*}
\centering
\includegraphics[width=0.55\hsize,clip]{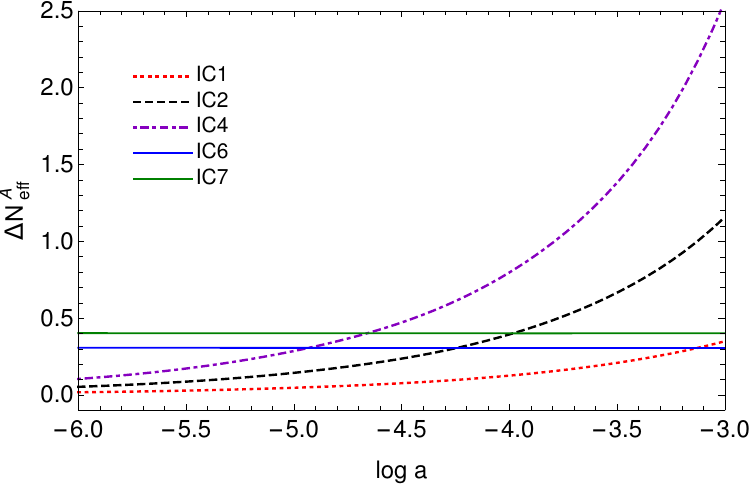}
\caption{Evolution of $\Delta N_{eff}^{A}$ versus the number of e-folds $N \equiv \log{a} = \alpha$ for the different initial conditions as specified before. Here, each of the $\beta_{i}=1$ solutions (solid curves) shows a constant contribution to $\Delta N_{eff}^{A}$ unlike the $\beta_{i}\neq1$ case.
} \label{fig:4}
\end{figure*}

Increasing the number of relativistic degrees of freedom in the early universe has been a promising programme in cosmology to ameliorate the Hubble tension \cite{Blinov:2020hmc}. This means that any extra contribution to radiation may affect the cosmic expansion rate during the BBN epoch (e.g. increasing the Hubble parameter at that period), the CMB anisotropies, and the LSS formation. We identify this kind of contribution to $N_{eff}$ as \textit{dark radiation}. Thus, the total relativistic energy density well after the electron-positron annihilation is given by
\begin{equation}
 \rho_{R}=\rho_{\gamma}+\rho_{\nu}+\rho_{A} \,,\label{eq39}
\end{equation}
where $\rho_{\nu}$ is the energy density of the standard active neutrinos and $\rho_{A}$ denotes the \textit{non-Abelian dark radiation}. The contribution $\Delta N_{eff}^{A}$ of these extra relativistic species can be estimated by comparing its contribution to the total energy density of the universe to the energy density in the CMB photons $\rho_{\gamma}$:
\begin{equation}
 \rho_{A}=\rho_{\gamma}\left[\frac{7}{8}\Delta N_{eff}^{A} \left(\frac{4}{11}\right)^{4/3}\right] \,,\label{eq40}
\end{equation}
such that 
\begin{equation}
\Delta N_{eff}^{A}(N)=\frac{\Omega_{A}(N)}{\chi\Omega_{\gamma}} \,.\label{eq41}
\end{equation}
In the previous expression, $\chi=\frac{7}{8} \left(\frac{4}{11}\right)^{4/3}$ and the explicit time-dependence of $\Omega_{A}$ has been shown through the number of e-folds $N$.  
As the vector field does not behave completely as radiation (since $w_{A}=1/3$ is not fulfilled during the whole evolution at least for solutions with $\beta_{i}\neq1$), $\Delta N_{eff}^{A}$ is indeed a function of the redshift. This feature is depicted in Fig.~\ref{fig:4} for the two cases of interest. For $\beta_{i}\neq1$,  $\Delta N_{eff}^{A}$ has almost a negligible effect during  BBN around $N\approx-6$ (see dotted-red, dashed-black and dot-dashed purple curves), leaving nucleosynthesis practically unaffected while at the time of recombination (around $N\approx-3.5$) the contribution is, in some cases (see e.g. the dashed-black and the dot-dashed purple curves), notoriously larger than the standard prediction. For $\beta_{i}=1$ the vector field behaves like radiation during all the cosmic evolution (see solid curves), thus contributing constantly to $\Delta N_{eff}^{A}$. The time-dependence of $\Delta N_{eff}^{A}$ is also enjoyed, for instance, by time-varying dark energy models like the one described in \cite{Lorenz:2017fgo}, and provides, in turn, subtle differences that can be used to constrain and characterize the effects of the vector field at different cosmic epochs in comparison with other phenomenological effects.

It is interesting to note that the upper limit $\Delta N_{eff}<0.30$ at 95$\%$ from the recent Planck release at recombination time \cite{Aghanim:2018eyx} leaves only the IC1 (with $\beta_{i}\neq1$) and IC6 (with $\beta_{i}=1$) solutions as viable options. As we see, due to the time dependence of $\Delta N_{eff}^{A}$, the non-Abelian dark radiation could leave distinct signatures on the BBN, CMB and LSS at different epochs of the universe's evolution such that combined analysis of different data set must be done carefully (see e.g. \cite{Schoneberg:2019wmt} for a similar suggested treatment). This implies that the above upper bound, and its derived constraints, can be different with the inclusion of new observational data. We illustrate better this point as follows: it is well known, for instance, that CMB data along with BAO measurements put tight constraints on $\Delta N_{eff}$ as imposed above, leading to a small window to extra relativistic contributions: $N_{eff}=2.99\pm0.17$ \cite{Aghanim:2018eyx} with $\Delta N_{eff}<0.30$. In the case of the the concordance cosmological model, its prediction $N_{eff} = 3.046$ is in good agreement with the above constraint. Interestingly, when measurements of the local Hubble parameter are included, $N_{eff} = 3.27\pm 0.15$ ($68\%$ C.L.) leading to a larger $\Delta N_{eff}\approx 0.4-0.5$ \cite{Blinov:2020hmc,Bernal:2016gxb}\footnote{Another possibility to increase $N_{eff}$ is when photons or electron-positron pairs are heated by the late decoupling of dark radiation, this is, after the time when the active neutrinos decouple \cite{Steigman:2013yua,Berlin:2019pbq}.}.
%

\subsection{Comparison with previous works}

Models of dark energy coupled to other dark sectors, such as neutrinos, offer a different avenue to ameliorate the tensions in the standard cosmological model. For instance, the coupling of neutrino (masses) to the quintessence field leaves important imprints on the anisotropies in the CMB \cite{Brookfield:2005bz,Brookfield:2005td} and can also relax previous constraints on the amount of early dark energy \cite{NobleChamings:2019ody}. On the other hand, most of the existing models of dark radiation
allow for interactions either between dark matter or different neutrino flavours at the time of recombination \cite{Blinov:2020hmc}. In this regard, our proposal is different since we are setting a new interface between dark radiation (represented by the Yang-Mills field) and the quintessence field. The idea of coupling the scalar field to both Abelian and non-Abelian vector fields has been investigated, however, early in the last decade but in the context of inflationary cosmology  \cite{Watanabe:2009ct,Murata:2011wv}. On the other hand, an interesting proposal in the context of dark energy scenarios allows interactions between the Higgs field and
the gauge vector field \cite{Rinaldi:2015iza,Orjuela-Quintana:2020klr}, leading to anisotropic scaling solutions with non-vanishing gauge fields similar to the present model. It is worth mentioning that the Higgs field and the vector gauge field are considered as a single dark energy fluid. Perhaps, a very similar approach to the one proposed here was introduced in \cite{Thorsrud:2012mu} where the scalar field is coupled both to matter and a
Maxwell-type vector field. The matter coupling is framed in the context of conformal transformations but the coupling of the Maxwell field to the scalar field is however modulated by a kinetic coupling function similar to the aforementioned anisotropic inflationary models. The couplings between the scalar field and vector fields in all these models, including the present one, have however a common feature: they generate the anisotropy in the expansion rate among other appealing characteristic effects of each model.

\section{Conclusions}
\label{conclusions}
There is strong evidence that new physics beyond the standard model of particle physics is required, for instance, to describe the presence of a still elusive dark matter component in the universe (see e.g.  \cite{Bertone:2016nfn}). There could exist, however, other dark sectors uncoupled from the standard model with unique and testable predictions that make them very attractive in search of explaining consistently the origin and evolution of the universe. In particular, phenomenological motivations come from discrepancies between early and late-time cosmological measurements, the lack of identification of the dark matter nature, the underlying mechanism of inflation, and the consensual explanation for the current accelerated expansion. The dark sectors that account for most of the energy content of the universe (dark energy and dark matter) can have presumably different geometrical structures that may be related each other, or even with other dark sectors as dark radiation, by a disformal transformation of the type in Eqn.~(\ref{eq3}).

The inclusion of new degrees of freedoms to the total energy budget of the universe carries with itself, generally, a significant cost and must be done with some caution since it may impact the cosmic history in an inconsistent way. We have investigated this premise assuming a dark radiation component consisting of a SU(2) gauge field disformally coupled to a scalar field (dark energy) but uncoupled from the standard matter sector. This approach clearly does not follow the idea behind traditional coupled dark energy models though, in essence, it can be treated similarly. According to this formulation, we have derived the field equations that show non-trivial derivative couplings of the scalar field to the vector field which translate into a non-vanishing vector field contribution during different stages in the evolution of the universe as obtained by a dynamical system analysis.  

Although it is not possible to study the global properties of the entire phase space as discussed, we have studied three appealing cases of cosmological interest: the uncoupled case with $\beta=1$, the isotropic case with $\Sigma=0$, and the Abelian case, the latter two with $\beta$ free. For these latter cases we have found new scaling solutions in terms of the disformal coupling parameters $\beta$ and $\lambda_{h}$ that can be interesting for coupled scenarios of the early universe. We have also found the standard quintessence fixed points as derived branches. Interestingly, the stability conditions and even the dynamical character of such points depend now on the coupling parameter $\lambda_{h}$. As a particular feature, most of the solutions found there belong to some branches of the uncoupled limit, exhibiting thus $\beta=1$ as a common fixed point among these solutions. In this regard, we have explored significantly most of the entire phase space by accessing some significant regions of it through the aforementioned cases.

Interestingly, we have found that the vector field equation of state acquires the novel form $w_{A}=\beta^{2}(N)/3$. Though $\beta=1$ is a stable attractor of the system, the vector field may experience a transient matter-radiation-like behaviour during the evolution of the universe before reaching its attractive-type character in the future. The main features of the model as well as the role of the vector field in the evolution of the cosmological background have been investigated by means of a numerical analysis. We conclude that, mostly depending on the initial conditions and on the $\beta$ parameter, the vector field can contribute notoriously to the energy content either as dark radiation (with $\beta_{i}=1$ and $w_{A}\approx1/3$) or as dark matter during the early universe (with $\beta_{i}\neq1$ and $w_{A}\approx0$), and making distinguishable the late-time dynamics with a (positive or negative) non-vanishing shear in comparison to the reference $\phi$CDM model. So, the universe inherits an anisotropic hair due to the presence of the non-Abelian gauge field at the present time in accordance with current bounds but losing it in the future. This is particularly interesting because the anisotropic expansion can contribute to the CMB quadrupole anomaly \cite{Appleby:2009za,Koivisto:2008ig} and thus
address the so-called quadrupole temperature problem \cite{Aghanim:2018eyx}.

Furthermore, small but visible changes are appreciated in the matter-radiation equality redshift that can be important in parameter estimation analysis. Finally, we estimated numerically the contribution of the non-Abelian dark radiation to the effective number of neutrinos which resulted intriguingly in a time-dependent function of the dark-to-proton density ratio (see Eqn.~(\ref{eq41})) with an almost negligible effect during BBN epoch but with a notorious contribution around recombination. It means that its effect may be traceable and constrained at different epochs of the universe's evolution.

In general, the novel anisotropic vector-scaling solutions found leave some imprints on the background cosmology either in the form of dark radiation or dark matter that may be tested with observations. So, we plan on constraining the phenomenological consequences observed in this model in a future work with the implementation of cosmological data at different redshifts. One of the phenomenological implications of this model is that the gauge field not only supports a preferred spatial direction, which may be relevant for the CMB quadrupole anomaly, but also can increase the energy density of dark matter and affects presumably the growth of structures in the universe.

On the other hand, there has been intense debate about the implications of the presence of dark radiation in the early universe on the Hubble tension, but without a consensual conclusion. We expect us to contribute to this issue by computing how extra relativistic species due to non-Abelian dark radiation may impact the Hubble parameter when different data set are considered, in particular measurements of the local Hubble parameter. Perhaps a more promising possibility to address the Hubble tension is to allow a small degree of anisotropy in the cosmic expansion rate of the late universe, by assuming that the large-scale spatial geometry of our universe is plane-symmetric, as was recently discussed in \cite{Akarsu:2019pwn,Cea:2022mtf}. Thus, non-vanishing contributions of the cosmic shear to the expansion history of the universe can contribute significantly to the universe expansion rate, leading to systematically larger values of $H_{0}$ compared to the standard $\Lambda$CDM model, but with a rather low statistical significance \cite{Akarsu:2019pwn} . The same conclusion was recently attained for the ellipsoidal universe, within about two standard deviations, by a simple examination of the comoving angular diameter distance which is linked to the Hubble parameter today \cite{Cea:2022mtf}.  It is important to mention that these results have been obtained by assuming a plain extension of the $\Lambda$CDM model with no underlying physical mechanism for generating the anisotropy in the expansion rate. Hence, it is very encouraging to investigate within our model whether the cosmic shear, sourced by the disformally coupled Yang-Mills field, can reconcile the higher value of the Hubble parameter today reported by the SH0ES team with the one reported by the Planck collaboration.

The impact of the vector field in the formation and evolution of the large scale structures is also quite challenging and has to be investigated at the perturbative level aiming at describing consistently the cosmic growth to reduce the tension of the matter density contrast between high and low-redshift measurements \cite{Macaulay:2013swa,Abbott:2017wau}. Whether these problems can be solved simultaneously or not in this model, it is an aspect that must be treated in the future with the use of cosmological data. So this is a long programme we have to address to assess the cosmological viability of this model.

At the theoretical level, the inclusion of new degrees of freedom, as proposed here, can also be framed in more general scalar-tensor (for some reviews see \cite{Kobayashi:2019hrl,Deffayet:2013lga,Heisenberg:2018vsk,Rodriguez:2017ckc}), vector-tensor \cite{GallegoCadavid:2021par,GallegoCadavid:2020dho,Jimenez:2019hpl,Gomez:2019tbj,GallegoCadavid:2019zke,Kase:2018tsb,Papadopoulos:2017xxx,Kimura:2016rzw,Jimenez:2016upj,Heisenberg:2016eld,Allys:2016kbq,Allys:2016jaq,Jimenez:2016isa,Allys:2015sht,Heisenberg:2014rta,Tasinato:2014eka}, and scalar-vector-tensor theories \cite{Heisenberg:2018acv}, but it will affect, in an unknown manner, the conditions for the avoidance of kinetic and Laplacian instabilities of their respective scalar and vector perturbations. Hence, there is no guarantee that the inclusion of a new dark sector leaves the theory free of pathologies. For instance, the mere presence of a SU(2) gauge field in the non-Abelian case contributes to additional tensor modes. It means that the tensor sector is now expressed in terms of four dynamical modes associated to the polarization states of the metric and the gauge field \cite{Namba:2013kia,Gomez:2019tbj}.
The presence of these additional tensor modes can produce different observational signatures such as modulations of the waveform, oscillations of the gravitational wave luminosity distance, and an anomalous gravitational wave speed (see e.g. \cite{Jimenez:2019lrk}). It is also possible that  the propagation speed of gravitational waves in the Abelian case could be modified due to the derivative coupling of the scalar field to the gauge field similar to what happens in scalar-vector-tensor theories \cite{Heisenberg:2018mxx}. Hence, we expect that further constraints on the parameter space may be placed from preventing instabilities in different (scalar, vector, and tensor) sectors of the model such as no-ghost conditions and non-trivial modifications to the propagation speeds. This is a relevant aspect that deserves to be investigated in order to guarantee the absence of any theoretical pathology and, therefore, the viability of the model.

\acknowledgments

L. G. G acknowledges financial support from Vicerrector\'ia de Investigaci\'on, Desarrollo e Innovaci\'on - Universidad de Santiago de Chile, project DICYT,
code 042031CM$\_$POSTDOC. Y. R. and J. P. B. A. are financed by the Patrimonio Aut\'onomo - Fondo Nacional de Financiamiento para la Ciencia, la Tecnolog\'ia y la Innovaci\'on Francisco Jos\'e de Caldas (MINCIENCIAS - COLOMBIA) Grant No. 110685269447 RC-80740-465-2020, project 69553.  Y. R. has received funding/support from the European Union's Horizon 2020 research and innovation programme under the Marie Sklodowska -Curie grant agreement No 860881-HIDDeN. 

\appendix
\section{Stability conditions for the uncoupled case} \label{app}

We show in this appendix the conditions for ensuring the stability of the fixed points and, thus, their dynamical character, consistent with the  conditions already found for their existence. Finding out these conditions cannot be done, however, in a compact way as was presented in sec.~\ref{subsec:4.1}, but we attempt to keep the generality as much as possible.

The eigenvalues associated to the radiation point are
\begin{equation}
    \left\{0,0,0,1,\frac{3}{2},\frac{-\lambda_{V}^{2}\pm\sqrt{3}\sqrt{32\lambda_{V}^{2}-5\lambda_{V}^{4}}}{2\lambda_{V}^{2}},\frac{-\lambda_{V}^{2}\pm\sqrt{\lambda_{V}^{4}-24q^{2}\lambda_{V}^{4}-48p^{2}s^{2}\lambda_{V}^{4}}}{2\lambda_{V}^{2}}\right\} \,. \label{App:eq1}
\end{equation}
From here, it is straightforward to see that the $(-)$ branches of the eigenvalues have always negative real parts for any value of $\lambda_{V}$ with the exception of $\lambda_{V}\neq0$. Hence, the real parts of some eigenvalues have opposite signs pointing out its saddle-type nature.
Let us discuss a little about the stability of some suitable submanifolds compatible with the radiation era. As we discussed, the radiation point supports solutions with $x=y=0$ whose eigenvalues are %
 \begin{equation}
\left\{2, \frac{3}{2}, -1, 1, 0, 0, 0, \frac{1}{2} (-1\pm\sqrt{1 - 24 q^{2} - 48 p^{2} s^{2}})\right\} \,.
 \end{equation}
%
It is easy to see also that the branches $s=0$ with $x\neq0$ and $y\neq0$, and $x=y=0$, provide similar eigenvalues, however, the latter cannot be contained within Eqn.~(\ref{App:eq1}).

Regarding the general matter domination solution, its eigenvalues point out to a saddle behaviour according to the standard criterion: 
\begin{equation}
\left\{0, -1, -\frac{3}{4}, -\frac{3}{4}, \frac{5}{4}, \frac{3\left(-2\lambda_{V}^{8}\pm \sqrt{2}\sqrt{36\lambda_{V}^{14}- 3\lambda_{V}^{16}\pm\sqrt{1296\lambda_{V}^{28} - 744\lambda_{V}^{30}+121\lambda_{V}^{32}}}\right)}{8\lambda_{V}^{8}}\right\} \,.
\end{equation}
For the branch $x=y=0$, the eigenvalues take the simple form
\begin{equation}
 \left\{-\frac{9}{4}, -\frac{3}{2}, \frac{3}{2}, \frac{5}{4}, -1, -\frac{3}{4}, -\frac{3}{4}, \frac{3}{4}, 0\right\} \,.   
\end{equation}
In the absence of the vector field, there are two possibilities: the one with $x=y=0$, which corresponds to the standard fully radiation dominated solution, with eigenvalues
\begin{equation}
 \left\{-\frac{3}{2}, -\frac{3}{2}, \frac{3}{2}, \frac{5}{4}, -1, -\frac{1}{2}, -\frac{1}{2}, -\frac{1}{4}, -\frac{1}{4}\right\} \,, 
\end{equation}
and the other one with $x=\pm y\neq0$ with
\begin{equation}
 \left\{-\frac{3}{2}, -1, -\frac{1}{2}, -\frac{1}{2}, -\frac{1}{4}, -\frac{1}{4}, \frac{5}{4},\frac{3(-\lambda_{V}^{3}\pm\sqrt{36\lambda_{V}^{4}-7\lambda_{V}^{6}})}{4\lambda_{V}^{3}}\right\}\,.
\end{equation}
Despite the fact that all branches have associated different eigenvalues, they all have at least one positive and one negative eigenvalue so their dynamical character is preserved.

Finally, the last fixed point corresponds to dark energy dominance with eigenvalues given by
\begin{equation}
\left\{\frac{\lambda_{1}}{3},\frac{\lambda_{1}}{3},\frac{\lambda_{2}}{6},\frac{\lambda_{2}}{6},\frac{\lambda_{2}}{3},\frac{\lambda_{2}}{3},\frac{2\lambda_{2}}{3},\frac{\lambda_{3}}{6},\frac{\lambda_{4}}{3}\right\} \,,
\end{equation}
with $\lambda_{1}=-9+\lambda_{V}^{2}$, $\lambda_{2}=-6+\lambda_{V}^{2}$, $\lambda_{3}=3+\lambda_{V}^{2}$, and $\lambda_{4}=-9+2\lambda_{V}^{2}$. At first glance, one can classify this point as a saddle point since it has, for any value of $\lambda_{V}$, one positive eigenvalue, this is, the one written in terms of $\lambda_{3}$. Nonetheless, when one looks at its associated eigenvector, it is a repeller only in the $\dbtilde{g}$-direction in phase space but, since this variable is not certainly a physical one and all the other variables are attracted to their corresponding values in the point, we say that this fixed point is an attractor point. Thus, demanding $\lambda_{1},\lambda_{2},\lambda_{4}<0$, we get $-\frac{3}{\sqrt{2}}<\lambda_{V}<\frac{3}{\sqrt{2}}$ which contains the condition for acceleration $-\sqrt{3}<\lambda_{V}<\sqrt{3}$. Hence, this point is a stable attractor describing the present accelerated expansion.


\section{Stability conditions for the isotropic case} \label{app1}

We start by reporting the eigenvalues for the fixed point eqn.~(\ref{eq28a})
\begin{equation}
    \left\{0,0,0,1, -\frac{4(\lambda_{h}+\lambda_{V})}{\lambda_{V}},\frac{-\lambda_{V}^{4}-\sqrt{3}\sqrt{32\lambda_{V}^{6}-5\lambda_{V}^{8}}}{2\lambda_{V}^{4}},\frac{-\lambda_{V}^{4}+\sqrt{3}\sqrt{32\lambda_{V}^{6}-5\lambda_{V}^{8}}}{2\lambda_{V}^{4}}\right\} \,. \label{App1:eq1}
\end{equation}
It is plain to see that it cannot be a repeller for any values of the model parameters. So, this is simply a saddle point. 

Although not discussed in the isotropic case, eigenvalues of the disformal and isotropic version of the scaling matter solution eqn.~(\ref{eq30}) are
\begin{equation}
    \left\{0,-1,-\frac{1}{2},-\frac{1}{4},-\frac{3(\lambda_{h}+\lambda_{V})}{\lambda_{V}},\frac{3\left(-\lambda_{V}^{6}-\sqrt{36\lambda_{V}^{10}-7\lambda_{V}^{12}}\;\right)}{4\lambda_{V}^{6}},\frac{3\left(-\lambda_{V}^{6}+\sqrt{36\lambda_{V}^{10}-7\lambda_{V}^{12}}\;\right)}{4\lambda_{V}^{6}}\right\} \,. \label{App1:eq2}
\end{equation}
Interestingly, it can be an attractor point for certain values of the model parameters. However, this corresponds to a matter dominated solution ($w_{eff}=0$) such that it cannot provide cosmic acceleration. Demanding, on the other hand, that at least one, either the fifth or the seventh, eigenvalue has positive sign leads, respectively, to  the constraints: $(\lambda_{h}<0\; \land \; 0<\lambda_{V}<-\lambda_{h}) \; \lor \; (\lambda_{h}>0 \; \land  -\lambda_{h}<\lambda_{V}<0)$ and $-\frac{3}{\sqrt{2}}<\lambda_{V}<0\; \lor \; 0<\lambda_{V}<\frac{3}{\sqrt{2}}$.

Eigenvalues for the attractor quintessence point are
\begin{equation}
    \left\{0,-\frac{2}{3}\lambda_{V}(\lambda_{h}+\lambda_{V}),\frac{1}{3}(\lambda_{V}^{2}-9),\frac{1}{3}(\lambda_{V}^{2}-6),\frac{1}{6}(\lambda_{V}^{2}-6),\frac{1}{3}(2\lambda_{V}^{2}-9),\frac{2}{3}(\lambda_{V}^{2}-6)\right\} \,. \label{App1:eq3}
\end{equation}
Demanding eigenvalues with real negative part, consistent with the condition for accelerated expansion, yields the following ranges: \\ $\left\{\lambda_{h}\leq-\sqrt{3} \, \land \, -\sqrt{3}<\lambda_{V}<0\right\}\quad \lor \quad \left\{-\sqrt{3}<\lambda_{h}\leq0\, \land \, (-\sqrt{3}<\lambda_{V}<0 \, \lor \, -\lambda_{h}<\lambda_{V}<\sqrt{3})\right\} \quad \lor \quad \left\{0<\lambda_{h}<\sqrt{3} \, \land \, (-\sqrt{3}<\lambda_{V}<-\lambda_{h} \, \lor \, 0<\lambda_{V}<\sqrt{3})\right\} \quad \lor \quad \left\{\lambda_{h}\geq\sqrt{3} \, \land \, 0<\lambda_{V}<\sqrt{3}\right\}$. The parameter space however keeps unaltered, though portioned, with respect to the standard quintessence case.

Finally, the novel coupling vector-scalar radiation fixed point has, after some algebraic manipulations, the eigenvalues
\begin{equation}
    \left\{0,0,0,1,-\frac{1}{2}+\Xi,-\frac{1}{2}-\Xi\right\} \,, \label{App1:eq4}
\end{equation}
where $\Xi\equiv \sqrt{\frac{9}{4}-\frac{8(3+\beta^{2})}{(1+\beta^{2})\lambda_{h}^{2}}}$. It is easy to check numerically that, for $0<\beta<1$ and any value of $\lambda_{h}$, the real part of $\Xi$ is essentially zero, hence the corresponding eigenvalues are negative and, therefore, the point can be classified as a saddle point.

\section{Stability conditions for the Abelian case} \label{app2}

Some stability conditions are shown for particular fixed points regarding the Abelian case as a complement to the ones found in the uncoupled case. The anisotropic kination point  eq.~(\ref{eqabel0}) has eigenvalues 
\begin{equation}
    \left\{0,0,2,3, \frac{\beta\lambda_{h}^{2}-4\beta\lambda_{h}\sqrt{-9+\lambda_{h}^{2}}}{2\beta\lambda_{h}^{2}},\frac{\beta\lambda_{h}^{2}+2\beta\lambda_{h}\sqrt{-9+\lambda_{h}^{2}}}{\beta\lambda_{h}^{2}},\frac{3(\beta\lambda_{h}^{2}+\beta\lambda_{h}\lambda_{V})}{\beta\lambda_{h}^{2}}\right\} \,. \label{App2:eq1}
\end{equation}
The fixed point can be repeller or a saddle point depending on the sign of the real parts of the eigenvalues. In the former case, the following condition must be satisfied: $-2\sqrt{3}<\lambda_{h}\leq -3\; \lor \; 3\leq\lambda_{h}<4\sqrt{\frac{3}{5}}$. Thus, for $\lambda_{h}>4\sqrt{\frac{3}{5}}$, the dynamical character of the fixed point changes from a repeller to a saddle point. In the above, we have restricted the solutions to those whose region of the parameter space fulfill  the condition of acceleration.

Eigenvalues associated to the radiation fixed point eq.~(\ref{eqabel1}) are 
\begin{equation}
    \left\{0,0,-1,1,-\frac{4(\lambda_{h}+\lambda_{V})}{\lambda_{V}},\frac{-\lambda_{V}^{2}-\sqrt{3}\sqrt{32\lambda_{V}^{2}-5\lambda_{V}^{4}}}{2\lambda_{V}^{2}},\frac{-\lambda_{V}^{2}+\sqrt{3}\sqrt{32\lambda_{V}^{2}-5\lambda_{V}^{4}}}{2\lambda_{V}^{2}}\right\} \,, \label{App2:eq1a}
\end{equation}
which is a saddle point for any value of $\lambda_{h}$ and $\lambda_{V}$.

Eigenvalues for the scaling matter-scalar solution eqn.~(\ref{eq30b}) are
\begin{equation}
    \left\{-\frac{3}{2},-1,-\frac{1}{2},-\frac{1}{4},-\frac{3(\lambda_{h}+\lambda_{V})}{\lambda_{V}},\frac{3(-\lambda_{V}^{6}-\sqrt{36\lambda_{V}^{10}-7\lambda_{V}^{12}})}{4\lambda_{V}^{6}},\frac{3(-\lambda_{V}^{6}+\sqrt{36\lambda_{V}^{10}-7\lambda_{V}^{12}})}{4\lambda_{V}^{6}}\right\} \,, \label{App2:eq1b}
\end{equation}
where the fifth and seventh eigenvalue can invert the sign within several ranges of values of $\lambda_{h}$ and $\lambda_{V}$. For instance, demanding a saddle-like nature, the simple conditions come separately from the fifth and seventh eigenvalue: $(\lambda_{h}<0 \; \land \; 0<\lambda_{V}<-\lambda_{h})\; \lor \; (\lambda_{h}>0 \; \land \; -\lambda_{h}<\lambda_{V}<0)$ and $-\frac{3}{\sqrt{2}}<\lambda_{V}<0 \; \lor \; 0<\lambda_{V}<\frac{3}{\sqrt{2}}$, respectively. 

We also report the renewed eigenvalues for the standard quintessence in terms of $\lambda_{V}$ and $\lambda_{h}$:
\begin{equation}
    \left\{-\frac{2}{3}\lambda_{V}(\lambda_{h}+\lambda_{V}),\frac{1}{3}(\lambda_{V}^{2}-9),\frac{1}{3}(\lambda_{V}^{2}-9),\frac{1}{6}(\lambda_{V}^{2}-6),\frac{1}{3}(\lambda_{V}^{2}-6),\frac{2}{3}(\lambda_{V}^{2}-6),\frac{1}{3}(2\lambda_{V}^{2}-9)\right\} \,. \label{App2:eq2}
\end{equation}
Demanding eigenvalues with real negative part, consistent with the condition for accelerated expansion, leads to the same allowed parameter space as the isotropic case.









\bibliographystyle{unsrt}
\bibliography{biblio}

\end{document}